\documentclass[floatfix,epsf, prb,twocolumn,amsmath,amssymb,showpacs,superscriptaddress]{revtex4}
\usepackage{graphicx}
\usepackage{amsmath}
\usepackage{amssymb}
\usepackage[T1]{fontenc}
\usepackage{color}

\usepackage{subfigure}

\begin{document}\setlength{\unitlength}{1mm}

\title{Kronig-Penney model on bilayer graphene: spectrum and transmission periodic
in the strength of the barriers}

\author{M.~Barbier}
\affiliation{Department of Physics, University of Antwerp, Groenenborgerlaan 171, B-2020 Antwerpen, Belgium\\}
\author{P.~Vasilopoulos}
\affiliation{Department of Physics, Concordia University, 7141 Sherbrooke Ouest, Montr\'eal, Quebec, Canada H4B 1R6\\}
\author{F.~M.~Peeters}
\affiliation{Department of Physics, University of Antwerp, Groenenborgerlaan 171, B-2020 Antwerpen, Belgium\\}

\newcommand{\ud}{\mathrm{d}}
\newcommand{\kvec}[2]{\begin{pmatrix} #1 \\ #2 \end{pmatrix} }
\newcommand{\kvecc}[4]{\begin{pmatrix} #1 \\ #2 \\ #3 \\ #4 \end{pmatrix} }
\newcommand{\kvecje}[2]{\bigl( \begin{smallmatrix} #1 \\ #2 \end{smallmatrix} \bigr)}
\newcommand{\rvecc}[4]{\begin{pmatrix}#1 & #2 & #3 & #4 \end{pmatrix} }
\newcommand{\rvec}[2]{\begin{pmatrix}#1 & #2 \end{pmatrix} }
\newcommand{\rvecje}[2]{(\begin{smallmatrix} #1 & #2 \end{smallmatrix})}
\newcommand{\matt}[4]{\begin{pmatrix} #1 & #2 \\ #3 & #4 \end{pmatrix} }
\newcommand{\mattje}[4]{\bigl( \begin{smallmatrix} #1 & #2 \\ #3 & #4 \end{smallmatrix} \bigr)}
\newcommand{\vgll}[2]{\left\{ \begin{array}{c} #1 \\ #2 \end{array} \right. }
\newcommand{\vglL}[2]{\begin{equation} \left\{ \begin{align} #1 \\ #2 \end{align} \end{equation} \right. }
\newcommand{\pd}[2]{\frac{\partial #1}{\partial #2}}
\newcommand{\pdd}[2]{\frac{{\partial}^2 #1}{{\partial #2}^2}}
\newcommand{\pdje}[1]{\partial_{#1}}
\newcommand{\ve}{\varepsilon}

\begin{abstract}
We show that the transmission through single and double $\delta$-function 
potential barriers of strength 
$P=VW_b/\hbar v_F $ in bilayer graphene is periodic in $P$ 
with period $\pi$. For a certain range of $P$ values we find states that are 
bound to the potential barrier and 
that run along the potential barrier. Similar periodic behaviour 
is found for the conductance. 
The spectrum of a  periodic succession of 
$\delta$-function barriers (Kronig-Penney model) in bilayer graphene is 
periodic in $P$ with period $2 \pi$. For $P$ smaller than a critical value 
$P_c$, the spectrum exhibits two Dirac 
points while for $P$ larger than $P_c$ an 
energy gap opens. These results are extended to the case of a superlattice of 
$\delta$-function barriers with $P$ alternating in sign between successive 
barriers; the corresponding spectrum is periodic in P with period $\pi$.
\end{abstract}

 \maketitle
\section{Introduction}\label{sec1}
Graphene, a one-atom thick layer of carbon atoms, has become a research 
attraction pole since its experimental discovery in 2004
\cite{geim1}. Since carriers in graphene behave like relativistic and chiral 
massless fermions with a linear-in-wave vector spectrum, 
many interesting features could be tested with this material such as 
the Klein paradox\cite{klein,kats}, which was recently observed\cite{kex}, the 
anomalous quantum Hall effect, etc., see Ref. \onlinecite{cast} for two recent 
reviews. The effort to realise 
this Klein tunnelling through a potential barrier also lead to 
other interesting features, such as resonant tunnelling through double barriers 
\cite{per2}. With the possibility to fabricate devices with single-layer 
graphene, bilayer graphene has also been extensively investigated 
and been shown to possess 
extraordinary electronic behaviour, such as a gapless spectrum, in the absence 
of bias, and chiral carriers\cite{mccannrev,kats}. Many of these nanostructures 
could be given another functionality if based on  bilayer instead of 
single-layer graphene.

The electronic band structure can be modified by the application of a periodic 
potential and/or magnetic barriers. Such 
superlattices (SLs) are commonly used to alter the band structure of 
nanomaterials. 
In single-layer graphene already a number of papers relate their work to the 
theoretical understanding of such periodic 
structures\cite{parkanisotropy,sha,nori,bai,barb1,snyman1,nasir2}. 
Much less experimental and theoretical work has been done on bilayer 
graphene\cite{bai,barb2}.

We will study 
the spectrum, the transmission, and the conductance of  bilayer graphene 
through an array of potential barriers 
using a simple model: the Kronig-Penney (KP) 
model\cite{nonrelkp}, i.e. a one-dimensional periodic succession of 
$\delta$-function barriers on bilayer graphene. 
The advantage of such a model system is that, 1) a lot can be done analytically, 
2) the system is clearly defined,
3) and it is possible to show a number of exact relations. The present research 
is also motivated by our recent findings for single-layer graphene\cite{barb3}, 
where very interesting and unexpected properties were found, for instance, that 
the transmission and energy spectrum are periodic in the strength of the 
$\delta$-function barriers. Surprisingly, we find that for bilayer graphene 
similar, but 
different, properties are found as function of the strength of the 
$\delta$-function potential barriers.
Due to the different electronic spectra close to the Dirac point, i.e., linear 
for graphene and quadratic for bilayer graphene, we find very different 
transmission probabilities 
through a finite number of barriers and 
 very different energy spectrum, for a superlattice of $\delta$-function
barriers, between  single-layer and  bilayer graphene. 

The paper is organised as follows. In Sec.  II we briefly present the
 formalism. In Sec.~III we give results for the transmission and
conductance through a single $\delta$-function barrier. 
We dedicate Sec. IV  to  bound states of a single $\delta$-function barrier and
 Sec.~V to those of two such barriers. 
In Sec.  VI we present the spectrum for the KP model 
and in Sec.~VII that for an extended  KP model by considering two 
$\delta$-function barriers with opposite strength in the unit cell. 
Finally, in Sec.~VIII 
we make a summary and concluding remarks. 
\section{Basic formalism}\label{sec2}
We describe the electronic structure of an infinitely large flat graphene 
bilayer by the continuum nearest-neighbour, tight-binding model and 
consider solutions with energy and wave vector close to the $K$ ($K'$) point. 
The corresponding four-band Hamiltonian and eigenstates $\Psi$ are
\begin{equation}\label{eq2_1}
	\mathcal{H} =  \begin{pmatrix}  V & v_F\pi & t_\perp & 0 \\ 
	v_F\pi^\dagger & V & 0 & 0 \\ t_\perp & 0 & V & v_F\pi^\dagger \\ 
	0 & 0 & v_F\pi & V \end{pmatrix}, \quad  \quad \psi = \begin{pmatrix} \psi_A \\
      \psi_B \\ \psi_{B'} \\ \psi_{A'}  \end{pmatrix}.
\end{equation}
with $\pi = p_x + i p_y$ ($p_{x,y} = - i \hbar \pdje{x,y}$) and $p$ the momentum 
operator. We apply one-dimensional potentials $V(x,y) = V(x)$ and consequently 
the wave function can be written as $\psi(x,y) = \psi(x) e^{i k_y y}$ with the 
momentum in the y-direction a constant of motion. 
Solving the time-independent Schr\"odinger equation $\mathcal{H} \psi = E \psi$ 
we obtain, for constant $V(x,y) = V$, the spectrum and the eigenstates. 
The latter are given by Eq. (\ref{app1_6}) in App. \ref{app1} and the spectrum 
by Eq. (\ref{app1_3}) 
\begin{equation}\label{eq2_2}
\begin{aligned}
	\ve & = u + 1/2 \pm \sqrt{1/4 + k^2},\\
	\ve & = u - 1/2 \pm \sqrt{1/4 + k^2},
\end{aligned}
\end{equation}
where we used the 
dimensionless variables, $\ve = E/t_\perp$, $u = V/t_\perp$, 
$x \rightarrow x t_\perp/\hbar v_F$, $k_y \rightarrow \hbar v_F k_y / t_\perp$ 
and $\ve' = \ve - u$; 
$v_F = 10^{6}$ m/s, and 
$t_\perp = 0.39$ eV expresses the coupling between the two layers.

For later purposes we also give the frequently used two-band Hamiltonian
\begin{equation}\label{eq2_3}
    \mathcal{H} =  - \frac{{v_F}^2}{t_\perp}\begin{pmatrix}  0 & {\pi^\dagger}^2 \\ \pi^2 &
    0
    \end{pmatrix} + V,
\end{equation}
and the corresponding spectrum
\begin{equation}\label{eq2_4}
   E-V = \pm ({v_F}^2 \hbar^2/t_\perp)(k_x^2 + k_y^2).
\end{equation}
As seen, there are qualitative differences between the two spectra 
(compare Eq.~(\ref{eq2_4}) with Eq.~(\ref{eq2_2})) that will be 
reflected in those for the transmission and conductance in some of the cases 
studied.
As the approximation of the two-band Hamiltonian is only valid for $ E \ll t_\perp$ \cite{mccann}, 
we can expect a qualitative difference with the four-band Hamiltonian 
if $|E| \gtrsim t_\perp $.
\section{Transmission through a $\delta$-function barrier}\label{sec3}
We assume $|E - V| < t_\perp$ outside the barrier such that 
we obtain one pair of localised and one pair of travelling eigenstates in the 
well regions 
characterised by wave vectors $\alpha$ and $\beta$, where $\alpha$ is real and 
$\beta$ imaginary, see App. \ref{app1}. Consider an incident wave with wave 
vector $\alpha$ from the left (normalised to unity);  part of it
will be reflected, with amplitude $r$, and part of it will be transmitted
with amplitude $t$. Then the transmission is $T = |t|^2$. Also, there 
are growing and decaying evanescent 
states near the barrier, with coefficients $e_g$ and 
$e_d$, respectively. The relation between the coefficients can be 
written in  the form 
\begin{equation}\label{eq3_1}
    \mathcal{N} \kvecc{t}{0}{e_d}{0} = \kvecc{1}{r}{0}{e_g}.
\end{equation}
This leads to a system of linear  equations that can be written in matrix form
\begin{equation}\label{eq3_2}
    \kvecc{1}{0}{0}{0} =
    \begin{pmatrix}  N_{11} &  0  & N_{13} & 0\\
                     N_{21} &  -1 & N_{23} & 0\\
                     N_{31} &  0  & N_{33} & 0\\
                     N_{41} &  0  & N_{43} & -1
    \end{pmatrix} \kvecc{t}{r}{e_d}{e_g},
\end{equation}
with $N_{ij}$  the coefficients of the transfer matrix $\mathcal{N}$.
Denoting the matrix in Eq. (\ref{eq3_2}) by $\mathcal{Q}$, 
we can evaluate the coefficients 
from $\rvecc{t,}{r,}{e_d,}{e_g,}^T = \mathcal{Q}^{-1}
\rvecc{1,}{0,}{0,}{0}^T$. As a result, to obtain the transmission 
amplitude $t$ it is sufficient to find the matrix element 
$(\mathcal{Q}^{-1})_{11} = [N_{11} - N_{13} N_{31}/N_{33}]^{-1}$.\\

\begin{figure}[ht]
  \begin{center}
	\includegraphics[height=3cm]{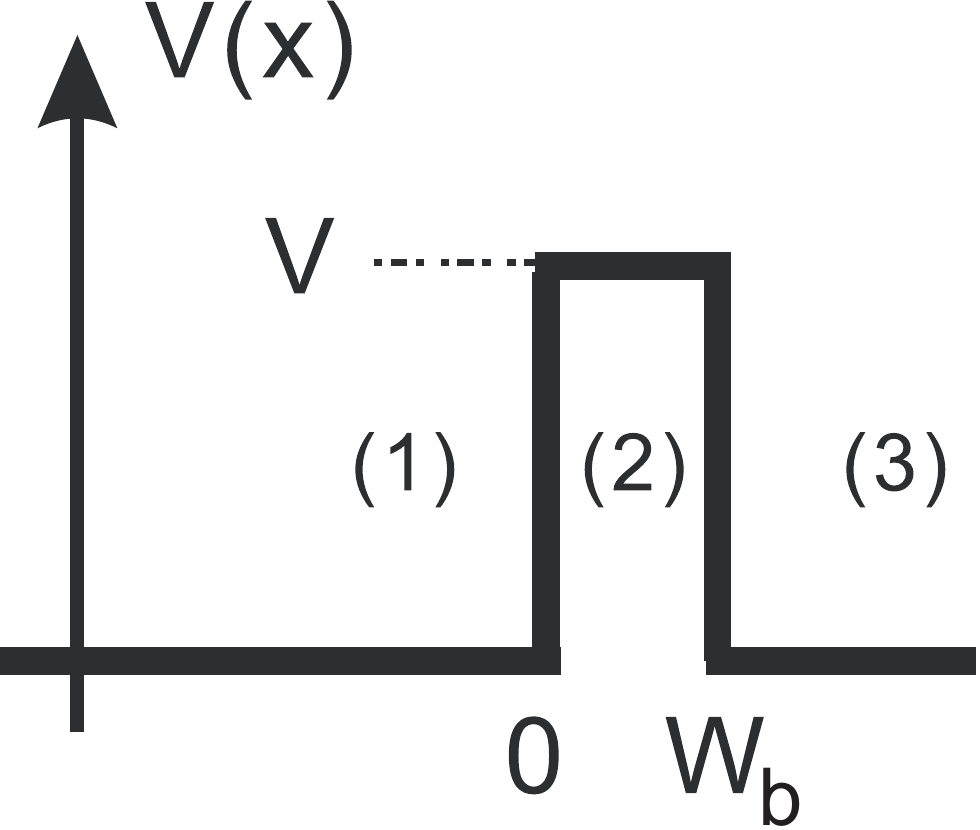}
    \end{center}
    \caption{Schematics of the potential V(x) 
    of a single square barrier.}\label{fig1}
\end{figure}

We model a $\delta$-function barrier as the limiting case of a square barrier, 
with height $V$ and width $W_b$ shown in Fig. \ref{fig1}, represented by 
$V(x) = V \Theta(x) \Theta(W_b - x)$.
The transfer matrix $N$ for this $\delta$-function barrier is calculated in 
appendix \ref{app2} and the limits $V \rightarrow \infty$ and 
$W_b \rightarrow 0$ are taken such that $P = V W_b / \hbar v_F$ is kept constant.

\begin{figure}[ht]
  \begin{center}
	\subfigure{\includegraphics[height=3.5cm]{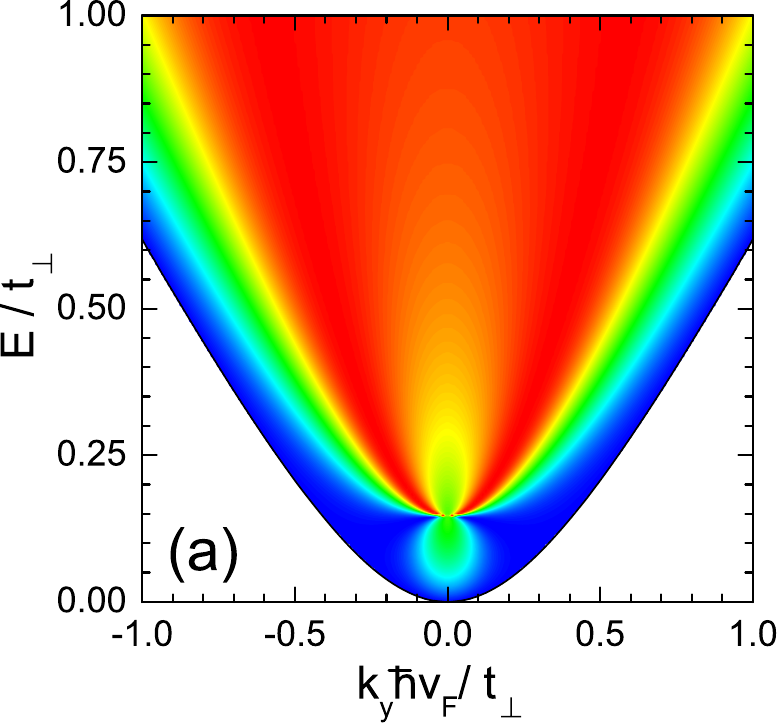}}
	\subfigure{\includegraphics[height=3.5cm]{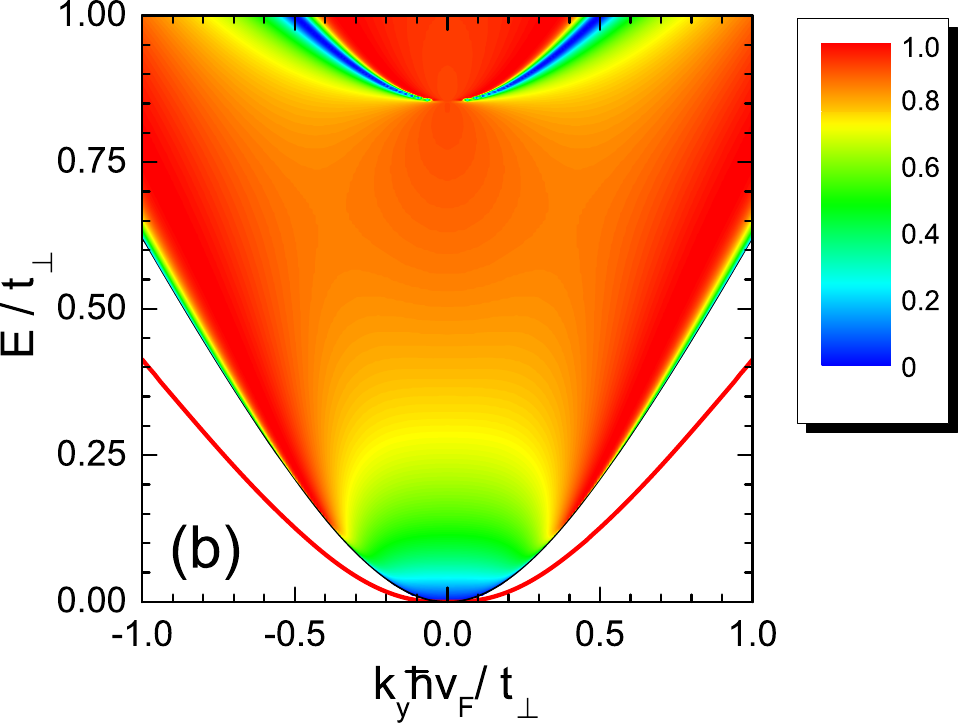}}
	\subfigure{\includegraphics[height=3.2cm]{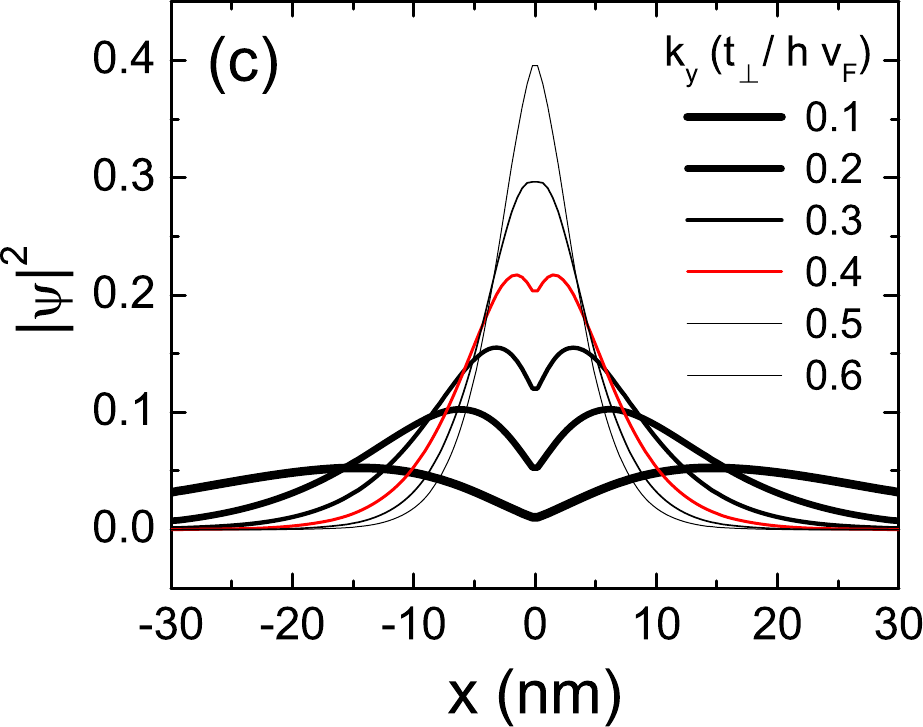}}
	\subfigure{\includegraphics[height=3.2cm]{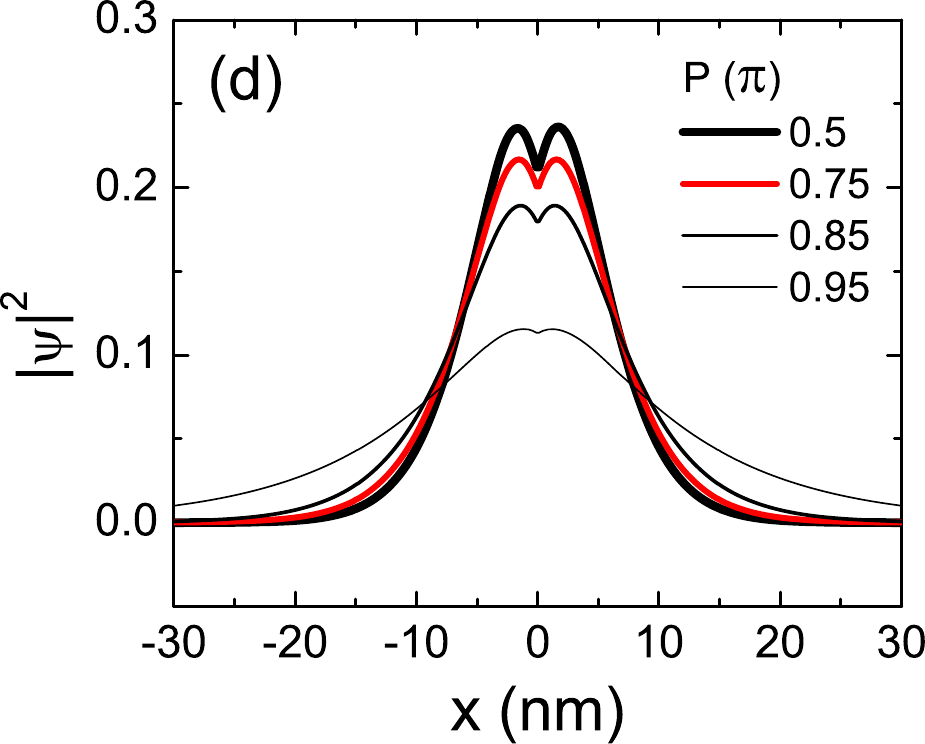}}
    \end{center}
  \caption{(Color online) Contour plot of the transmission for $P = 0.25\pi$ 
  in (a) and  $P = 0.75\pi$ in (b). 
  In (b) the bound state, shown by the red curve, is at positive energy. 
  The white area shows the part where $\alpha$ is imaginary. The probability 
  distribution $|\psi(x)|^2$ of the bound state is plotted 
  in (c) for various values of $k_y$ and in  (d) for different values of 
  $P$.}\label{fig2}
\end{figure}

The transmission $T = |t|^2$ for $\alpha$ real and $\beta$ imaginary 
is obtained from the inverse amplitude,
\begin{equation}\label{eq3_7}
	\frac{1}{t} = \cos P  + i\mu \sin P
	 + \frac{(\alpha - \beta)^2 k_y^2}{4 \alpha \beta \ve^2} \frac{\sin^2 P}{\cos P + i	\nu	\sin P}, 
\end{equation}
where $\mu=(\ve + 1/2)/\alpha$ and $\nu=(\ve -1/2)/\beta$. 
Contour plots of the transmission $T$ are shown in Figs. \ref{fig2}(a) and 
\ref{fig2}(b) for strengths $P = 0.25\pi$ and $P = 0.75 \pi$, respectively.

The transmission remains invariant under the transformations:
\begin{eqnarray}
\nonumber
\hspace*{-.59cm}& 1) & P\to P+n\pi , \\*
\hspace*{-.59cm}& 2) & k_y\to-k_y\,. 
\end{eqnarray}
The first property is in contrast with what is obtained in 
Ref. \onlinecite{kats}. 
In the latter work it was found, by using the $2\times 2$ Hamiltonian, that the 
transmission $T$ should be zero for $k_y \approx 0$ and $E < V_0$, while we 
can see here that for certain strengths $P = n\pi$ there is perfect 
transmission. 
The last property is due to the fact that $k_y$ only appears squared in the 
expression for the transmission. Notice that in contrast to single-layer 
graphene the transmission for $\ve \approx 0$ is practically zero. The cone for 
nonzero transmission shifts to $\ve = 1/2(1 - \cos P)$ with increasing $P$ till 
$P = \pi$. An area with $T = 0$ appears when $\alpha$ is 
imaginary, i.e., for  $\ve^2 + \ve - k_y^2 < 0$ (as no propagating states are 
available in this area, we expect bound states to appear). 
From Figs. \ref{fig2}(a), \ref{fig2}(b) it is apparent that the transmission in 
the forward direction, i.e., for $k_y \approx 0$, 
is in general smaller than $1$; 
accordingly, there is no Klein tunneling. However, for $P = n\pi$, with $n$ an 
integer, the barrier becomes 
perfectly transparent.

For $P = n \pi$ we have $V = \hbar v_F (n\pi/W_b)$. 
If the electron wave vector is $k = n \pi / W_b$ its energy equals the 
height of the potential barrier and consequently there is a quasi-bound state 
and thus a resonance \cite{matQdot}. The condition on the wave vector implies 
$W_b = n \lambda/2$ where $\lambda$ is the wavelength. This is 
the standard condition for Fabry-Perot resonances. 
Notice though that the invariance of the 
transmission under the change $P \rightarrow P + n\pi$ is not 
equivalent to the Fabry-Perot resonance condition.

From the transmission we can calculate the conductance $G$  given, at zero 
temperature, by
\begin{equation}\label{eq3_8}
	G/G_0 = \int_{-\pi/2}^{\pi/2} T(E,\phi) \cos\phi \ud \phi,
\end{equation}
where $G_0=(4 e^2/2 \pi h)[E_F^2 + t_\perp E_F]^{1/2}/\hbar v_F$; 
the angle of incidence $\phi$ is determined by $\tan \phi = k_y/\alpha$. 
It is not possible to obtain the conductance analytically, 
therefore we evaluate this integral numerically. 

\begin{figure}[ht]
  \begin{center}
	\subfigure{\includegraphics[height=3.4cm]{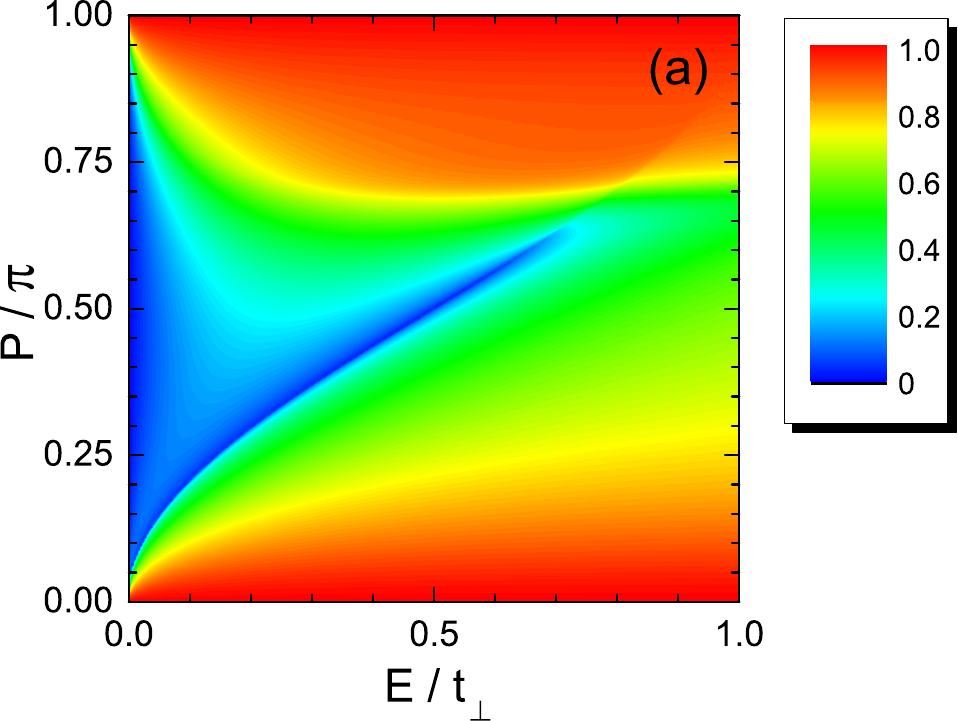}}
	\subfigure{\includegraphics[height=3.2cm]{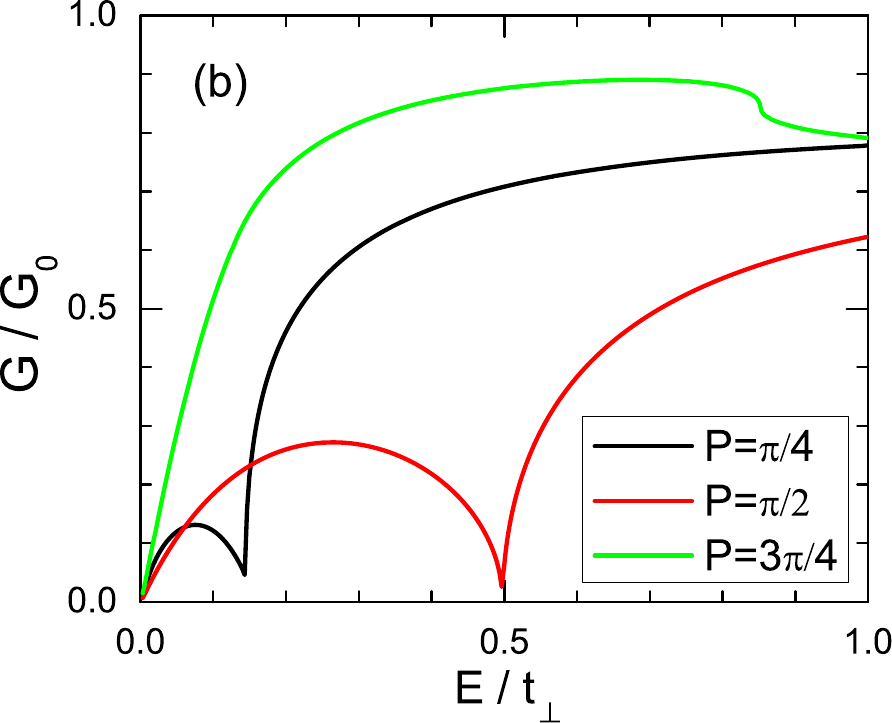}}
    \end{center}
	\caption{(Color online) (a) Contour plot of the conductance $G$.
	(b) Slices of $G$ along constant $P$.}\label{fig3}
\end{figure}

The conductance is a periodic function of $P$ (since the transmission is) 
with period $\pi$. 
Fig. \ref{fig3} shows a contour plot of the conductance for one period. 
As seen, 
the conductance has a sharp minimum at $\ve = 1/2(1 - \cos P)$:  this is due to 
the cone feature in the transmission which shifts to higher energies with 
increasing $P$. 
Such a sharp minimum was not present in the conductance of single-layer 
graphene when applied to the same $\delta$-function potential 
barrier\cite{barb3}.
\section{Bound states of a single $\delta$-function barrier}\label{sec4}
The bound states here are states that are localised in the x-direction close to 
the barrier but are free to move along the barrier, i.e. in the y-direction. 
Such bound states are characterised by the fact that the wave function decreases 
exponentially in the $x$ direction, 
i.e., the wave vectors $\alpha$ and $\beta$ are 
imaginary. This leads to 
\begin{equation}\label{eq4_1}
    \kvecc{e_{g1}}{0}{e_{g2}}{0} = \mathcal{N} \kvecc{0}{e_{d1}}{0}{e_{d2}},
\end{equation}
which we can write as
\begin{equation}\label{eq4_2}
    \kvecc{0}{0}{0}{0} = \mathcal{Q} \kvecc{e_{d1}}{e_{g1}}{e_{d2}}{e_{g2}},
\end{equation}
where the matrix $\mathcal{Q}$ is the same as in Eq. (\ref{eq3_2}).
In order for this homogeneous algebraic set of equations to have a nontrivial 
solution, the determinant of $\mathcal{Q}$ must be zero. 
This gives rise to a transcendental equation for the dispersion relation
\begin{equation}\label{eq4_3}
    \det \mathcal{Q} = N_{11} N_{33} - N_{13} N_{31} = 0,
\end{equation}
which can be written explicitly
\begin{equation}\label{eq4_4}
    [\cos P + i \mu\sin P ][\cos P + i \nu\sin P ]
  + \frac{(\alpha-\beta)^2 k_y^2}{4 \alpha \beta \ve^2}\sin^2 P = 0,
\end{equation}
This expression is invariant under the transformations
\begin{eqnarray}\label{eq4_5}
\nonumber
\hspace*{-.59cm}& 1) &  P \rightarrow P+n\pi, \\*
\nonumber
\hspace*{-.59cm}& 2) &  k_y \rightarrow -k_y, \\*
\hspace*{-.59cm}& 3) & (\ve,P) \rightarrow (-\ve,\pi - P)\, .
\end{eqnarray}
Furthermore, there is one bound state for $k_y > 0$ and  $\pi/2 < P < \pi$. For 
$P < \pi/2$ we can see that there is also a single bound state for negative 
energies from the third property above. 
From this transcendental 
formula one can find the solution for the energy $\ve$ as function of $k_y$ 
numerically. We show the bound state by the solid red curve 
in Fig. \ref{fig2}(b). This state is bound to the potential in the $x$ direction 
but moves as a free particle 
in the $y$ direction. We have two such states, one that moves along the $+y$ 
direction and one along the $-y$ direction. The numerical solution approximates 
the curve $\ve = \cos P\, [-1/2 + (1/4 + k_y^2)^{1/2}]$.
If one uses the $2\times2$ Hamiltonian one obtains the dispersion relation given 
in Appendix \ref{app3} by Eq. (\ref{app3_1}). By solving this equation 
one finds for each value of $P$ two bound states 
one for positive and one for negative $k_y$. 
Moreover,  for positive $P$ these bands have a hole like behaviour and for 
negative $P$ an electron like behaviour. Only for small $P$ do these results 
coincide with those from the $4\times4$ Hamiltonian.

The wave function $\psi(x)$ of such a bound state is characterised by the 
coefficients $e_{g1}$, $e_{g2}$ 
on the left, and $e_{d1}$ and $e_{d2}$ on the 
right side of the barrier. We can obtain the latter coefficients by using 
Eq. (\ref{eq4_2}), 
by assuming $e_{g1} = 1$ and afterwards  normalising the 
total probability to unity in dimensionless units. The wave function $\psi(x)$ 
to the left and right of the barrier can be determined from these coefficients 
by using 
Appendix \ref{app1}. 
In Figs. \ref{fig2}(c), (d) we show the probability distribution 
$|\psi(x)|^2$ 
of a bound state for a single $\delta$-function barrier:  in (c) we show it for 
several $k_y$ values and in (d) for different values of $P$. One can see that 
the bound state is localised around the barrier and is less smeared out with 
increasing $k_y$.  Notice that the bound state is more strongly confined 
for $P = \pi/2$ 
and that $|\psi(x)|^2$ is invariant under the transformation 
$P \rightarrow \pi - P$.
\section{Transmission through two $\delta$-function barriers}\label{sec5}
\begin{figure}[ht]
	\begin{center}
		\subfigure{\includegraphics[width=4cm]{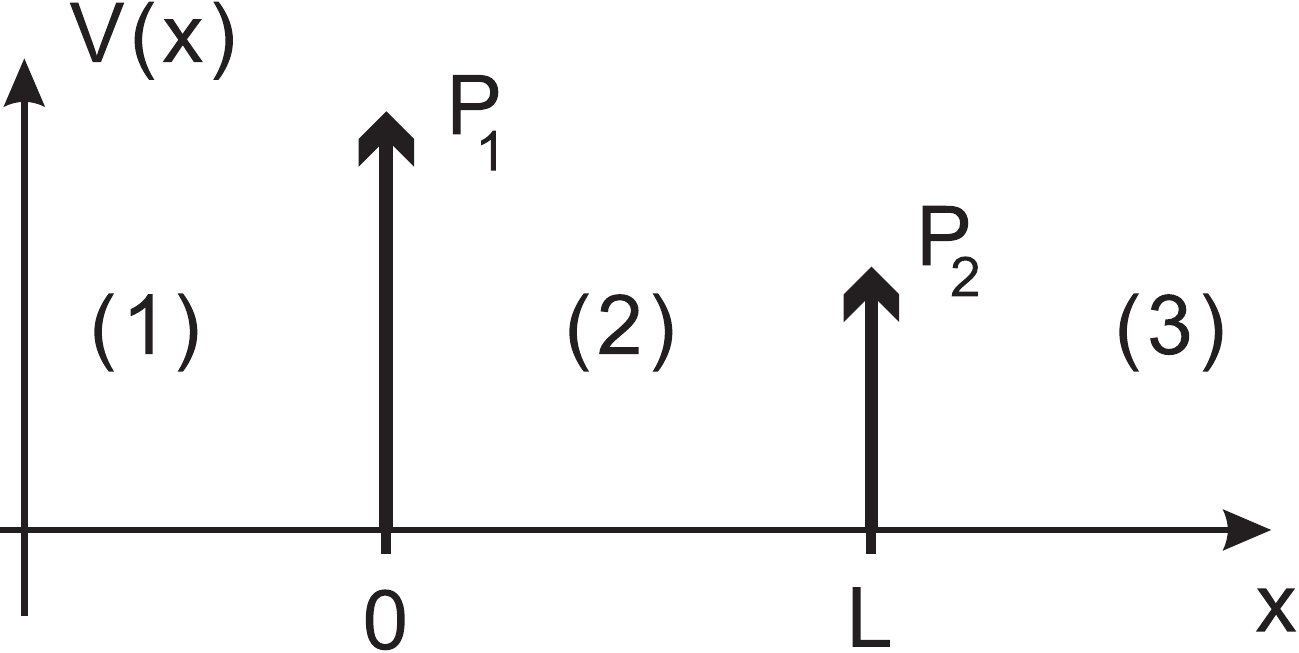}}
	\end{center}
	\caption{A system of two $\delta$-function barriers 
	with strengths $P_1$ and $P_2$ 
	placed a distance $L$ apart.}
	\label{fig4}
\end{figure}

We consider a system of two barriers, separated by a distance $L$, with 
strengths $P_1$ and $P_2$, respectively, 
as shown schematically in  Fig. \ref{fig4}. 
We have $L \rightarrow L t_\perp/\hbar v_F \equiv 0.59261 L/nm$ which for $L=10$ 
nm,  $v_F = 10^6$ m/s, and $t_\perp = 0.39$ eV equals $5.9261$ in dimensionless 
units. 
The wave functions in the different regions 
are related as follows
\begin{eqnarray}
\nonumber
	\psi_1(0) &=& \mathcal{S}_1\psi_{2}(0),\quad
	\psi_2(0) = \mathcal{S}'\psi_{2}(L),\\*
	\psi_{2}(1) &=& \mathcal{S}_2 \psi_{3}(L),\quad
	\psi_{1}(0) = \mathcal{S}_1 \mathcal{S}' \mathcal{S}_2 \psi_{3}(L),
\end{eqnarray}
where  $\mathcal{S}' = \mathcal{G} \mathcal{M}(1)\mathcal{G}^{-1}$ represents a 
shift from x=0 to x=L and the matrices $S_{1}$ and $S_{2}$ are equal to the 
matrix $\mathcal{N}'$ of Eq. (\ref{app2_6}) with $P = P_1$ and $P = P_2$, 
respectively.
Using the transfer matrix 
$\mathcal{N} = \mathcal{G}^{-1} \mathcal{S}_1 \mathcal{S}' \mathcal{S}_2 
\mathcal{G} \mathcal{M}(L)$ we obtain the transmission $T = |t|^2$. 
\begin{figure*}[ht]
  \begin{center}
	\subfigure{\includegraphics[height=3.2cm]{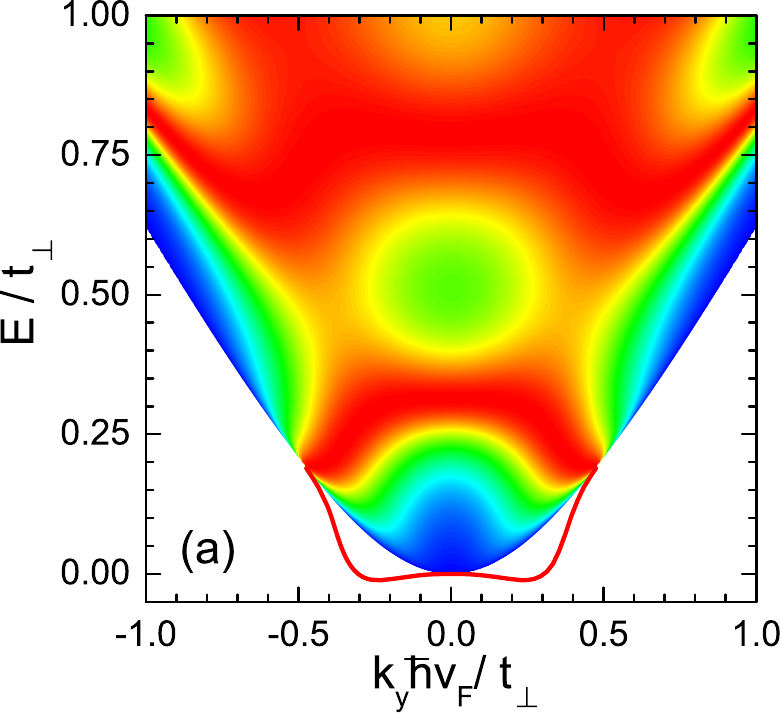}}
	\subfigure{\includegraphics[height=3.2cm]{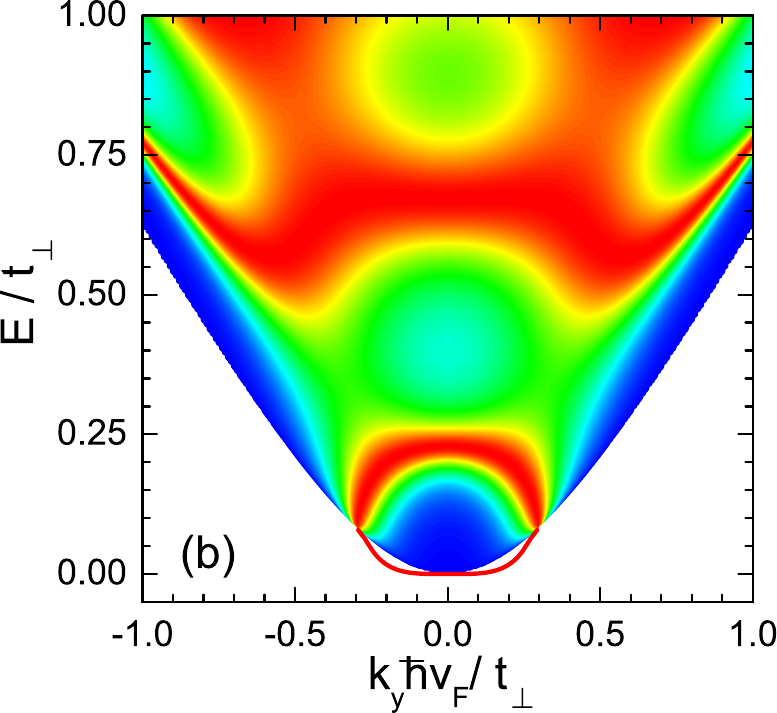}}
	\subfigure{\includegraphics[height=3.2cm]{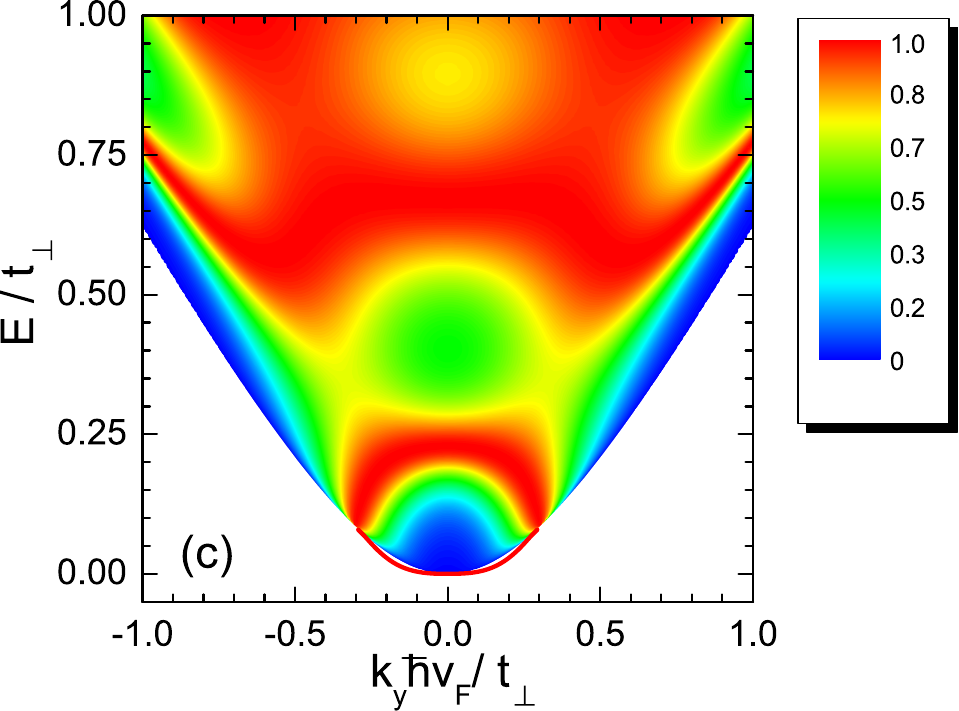}}
	\subfigure{\includegraphics[height=3.2cm]{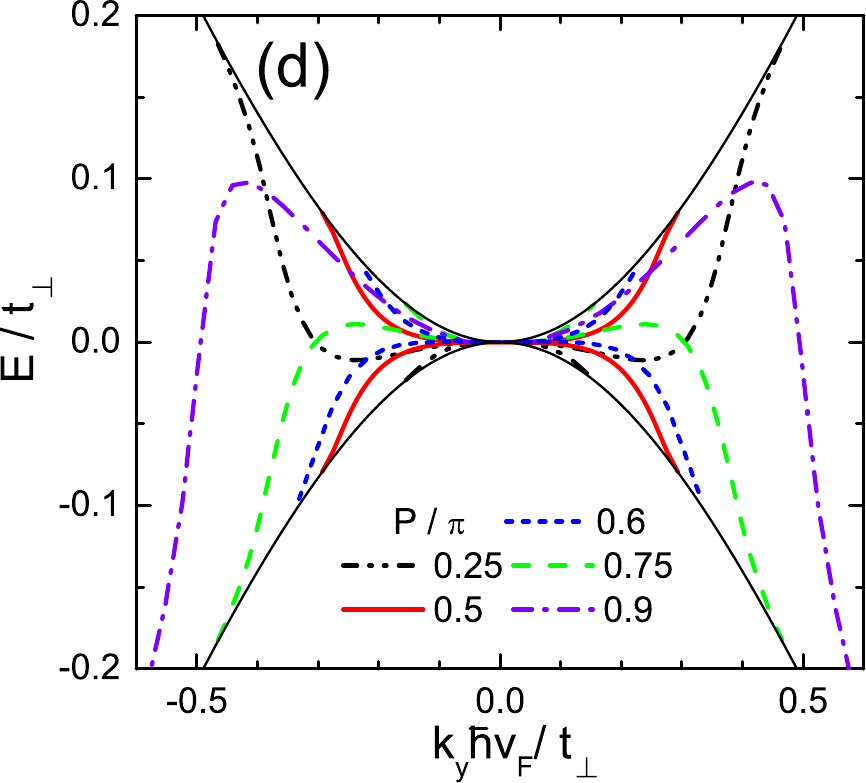}}
	\subfigure{\includegraphics[height=3.2cm]{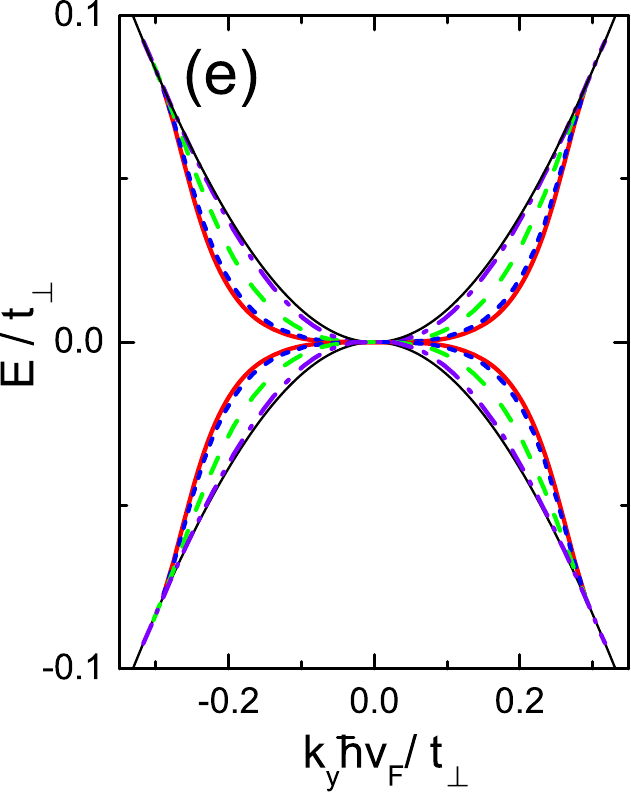}}
    \end{center}
  \caption{
  (Color online) Panels (a), (b), and (c): 
  contour plots of the transmission through two $\delta$-function barriers of 
  equal strength $P = |P_1| = |P_2|$ separated by a distance $L = 10$ nm. 
  For parallel barriers we took $P = 0.25\pi$ in (a) and $P = 0.5\pi$ in (b). 
  For anti-parallel barriers results are given for $P = 0.5\pi$ in (b) and $P = 0.25\pi$ in (c). 
  The solid red curves in the white background region is the spectrum for the bound states.
  Panels (d) and (e) show the dispersion relation of the bound states for various strengths $|P|$, 
  respectively, for parallel and anti-parallel barriers. The thin black curves 
  delimit the region where bound states are possible.
  }\label{fig5}
\end{figure*}

In Fig. \ref{fig5} the transmission $T(\ve,\,k_y)$ is shown for parallel 
(a), (b) and anti-parallel (b), (c) $\delta$-function barriers with equal 
strength, 
i.e., for  $|P_1| = |P_2|$, that are separated by $L = 10$ nm, with 
$P = 0.25 \pi$ in (a) 
and $P=0.5 \pi$ in (b). For $P = \pi/2$, the transmission amplitude $t$ for 
parallel 
barriers equals $-t$ for anti-parallel ones and the transmission $T$ is the 
same, as well  the formula for the bound states.  Hence panel (b) is the same 
for parallel and anti-parallel barriers. The contour plot of the transmission 
has a very particular structure which is very different from the single-barrier 
case. There are two bound states for each sign of $k_y$, 
which are shown in panel (d) for parallel and panel (e) for anti-parallel 
barriers. For anti-parallel barriers these states have mirror-symmetry with 
respect to $\ve = 0$ but for parallel barriers this symmetry is absent. 
For parallel barriers 
the change $P \rightarrow \pi - P$ 
will flip the spectrum of the bound state. 
The spectrum of the bound states extends into the low-energy transmission region
and gives rise to a pronounced resonance. 
Notice that for certain $P$ 
values (Figs. \ref{fig5}(a) and \ref{fig5}(d)) the energy 
dispersion for the bound state has 
a camelback shape for small $k_y$, indicating free propagating states along the 
$y$ 
direction with velocity  opposite to that 
for larger $k_y$ 
values. 
Contrasting Fig. \ref{fig2}(b) with Fig. \ref{fig5}(d)-(e)we see that the 
free-particle like spectrum of Fig. \ref{fig2}(b) for the bound 
states of a single $\delta$-function barrier is strongly modified when  
two $\delta$-function barriers are present.

From the transfer matrix we find that the 
transmission is invariant under 
the change $P \rightarrow P+n\pi$ and $k_y \rightarrow -k_y $ 
for parallel barriers,
which was also the case of a single barrier, cf. Eq. (\ref{eq4_5}).
In addition, it is also invariant,
for anti-parallel barriers, under the change
\begin{equation}
	P\to \pi-P.
\end{equation}

\begin{figure}[ht]
  \begin{center}
	\subfigure{\includegraphics[height=3.4cm]{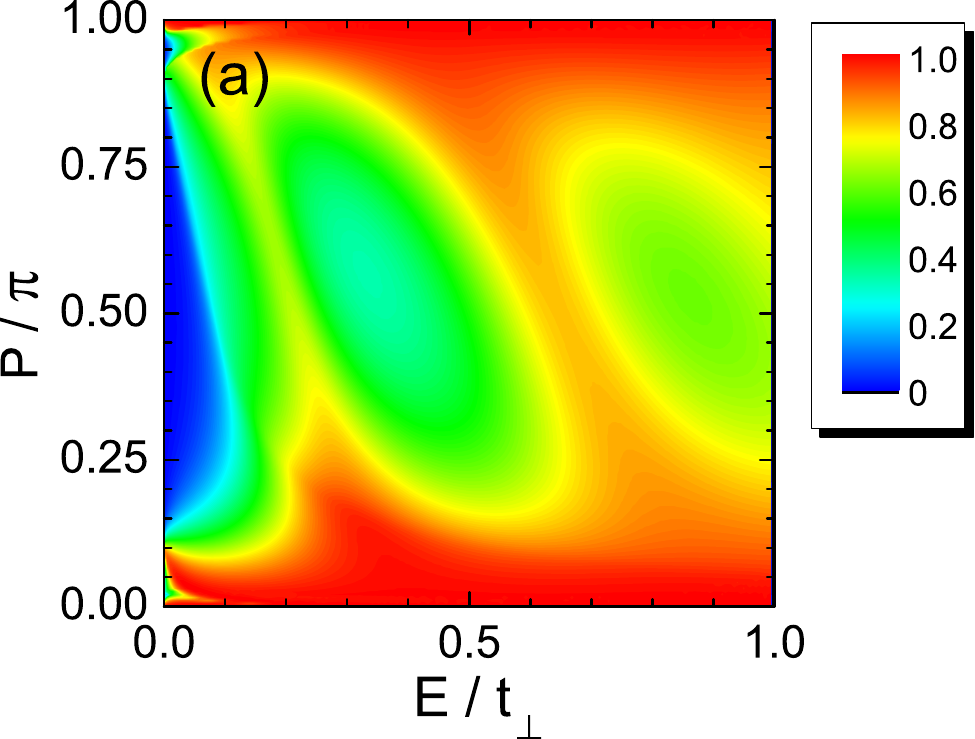}}
	\subfigure{\includegraphics[height=3.2cm]{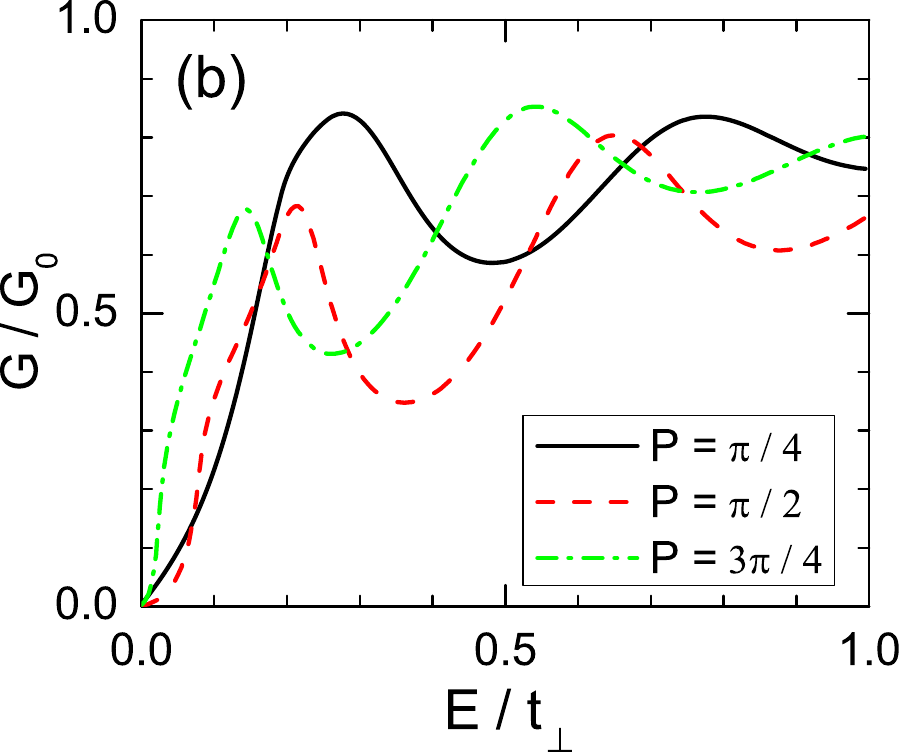}}\\
 	\subfigure{\includegraphics[height=3.4cm]{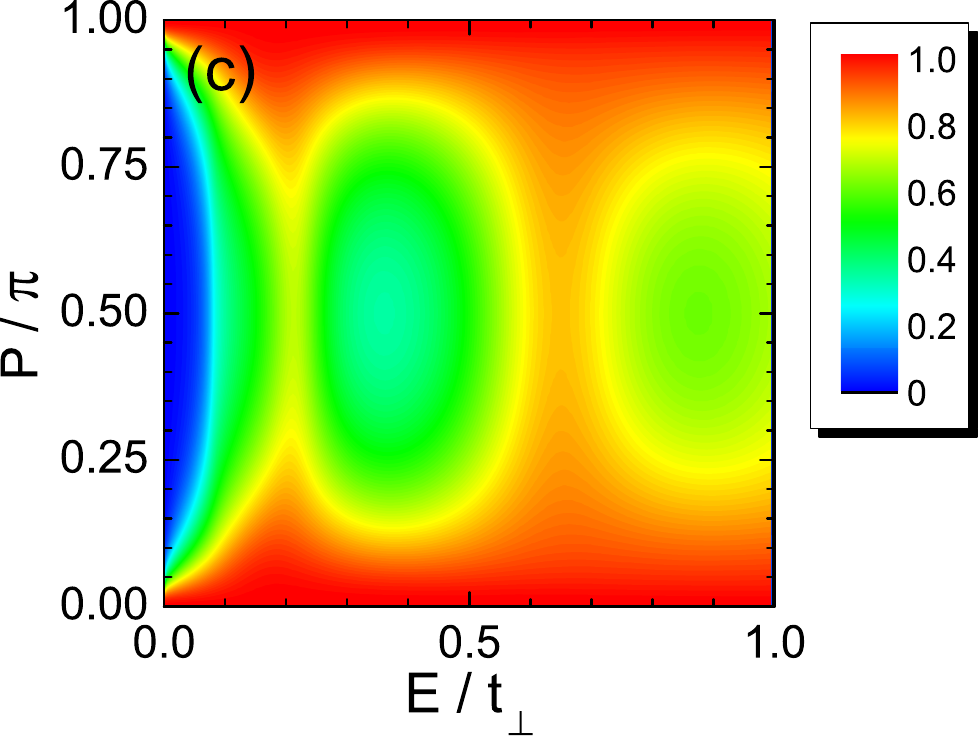}}
	\subfigure{\includegraphics[height=3.2cm]{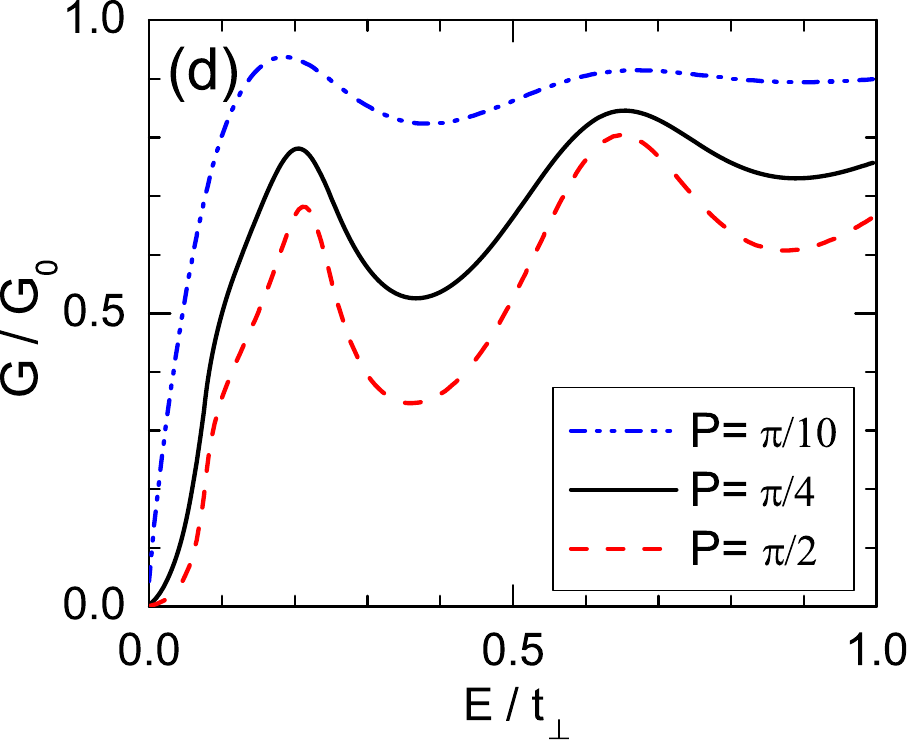}}
    \end{center}
  \caption{(Color online) Contour plot of the conductance 
  of two $\delta$-function barriers with strength $|P_2| = |P_1| = P$ and 
  inter-barrier distance $L = 10$ nm.  Panel (a) is for parallel barriers 
  and panel (c) for anti-parallel barriers. Panels (b) and (d) show 
  the conductance, along constant $P$, extracted from panels
  (a) and (c), respectively.}\label{fig6}
\end{figure}

The conductance $G$ is calculated  numerically 
as in the case of a single barrier. 
We show it 
for (anti-)parallel $\delta$-function 
barriers of equal strengths in Fig. \ref{fig6}. The symmetry 
$G(P + n\pi) = G(P)$ of the single barrier conductance holds here as well. 
Further, we see that for anti-parallel barriers $G$ has the additional symmetry 
$G(P) = G(\pi - P)$ as the transmission does.
\section{Kronig-Penney model}\label{sec6}
We consider an infinite sequence of 
equidistant
$\delta$-function potential barriers, i.e., a 
superlattice (SL), with potential
\begin{equation}\label{eq6_1}
	V(x) = P \sum_n \delta(x-n L).
\end{equation}
As this potential is periodic the wave function should be a Bloch function. 
Further, we know how to relate the coefficients $\mathcal{A}_1$ of the wave 
function before the barrier with those 
($\mathcal{A}_3$) after it, see Appendix \ref{app2}. 
The result is
\begin{equation}\label{eq6_2}
	\psi(L) = e^{i k_x L} \psi(0),\quad \mathcal{A}_1 = \mathcal{N} \mathcal{A}_3,
\end{equation}
with $k_x$ the Bloch wave vector. From these boundary conditions we can extract the relation
\begin{equation}\label{eq6_3}
	e^{-i k_x L} \mathcal{M}(L) \mathcal{A}_3 = \mathcal{N} \mathcal{A}_3,
\end{equation}
with the matrix $\mathcal{M}(x)$ given by Eq. (\ref{app1_5}). 
The determinant of the coefficients in Eq. (\ref{eq6_3}) must be zero, i.e.,
\begin{equation}\label{eq6_4}
	\det [e^{-i k_x L} \mathcal{M}(L) - \mathcal{N}] = 0.
\end{equation}

If $k_y = 0$, which corresponds to the pure 1D case, one can easily obtain the 
dispersion relation because the first two and the 
last two components of the 
wave function decouple. Two transcendental equations are found
\begin{subequations}\label{eq6_5}
\begin{eqnarray}
\hspace*{-1.2cm}&&\cos k_x L  = \cos \alpha L \cos P + \frac{1}{2}\left(\frac{\alpha}{\ve} + \frac{\ve}{\alpha}\right)\sin \alpha L \sin P ,\\*
\hspace*{-1.2cm}&&\cos k_x L  = \cos \beta L \cos P + \frac{1}{2}\left(\frac{\beta}{\ve} + \frac{\ve}{\beta}\right)\sin \beta L \sin P.
\end{eqnarray} 
\end{subequations}
Since $\beta$ is imaginary for $0 < E < t_\perp$, 
we can write Eq. (\ref{eq6_5}b) as
\begin{equation}\label{eq6_6}
\hspace*{-0.2cm}\cos k_x L  = \cosh |\beta| L \cos P - \frac{|\beta|^2+\ve^2}{2|\beta|\ve}\sinh |\beta| L \sin P ,
\end{equation}
which makes it easier to compare with the spectrum of the KP model obtained 
from the $2\times2$ Hamiltonian, see Eq. (3). The latter is given by the two relations
\begin{subequations}\label{eq6_7}
\begin{eqnarray} 
	\cos k_x L &=& \cos \kappa L + (P/2 \kappa) \sin \kappa L,\\
	\cos k_x L &=& \cosh \kappa L - (P/2 \kappa) \sinh \kappa L,
\end{eqnarray}
\end{subequations} 
with $\kappa = \sqrt{\ve}$. This dispersion relation, 
which 
has
the same form as the one for standard electrons, {\it  is not periodic in $P$} 
and the difference from that of the four-band Hamiltonian is due to the fact 
that the 
former is not valid for high potential barriers. 
One can also contrast the dispersion relations (\ref{eq6_5}) and (\ref{eq6_7}) 
with the corresponding one on single-layer graphene \cite{barb3}
\begin{equation} 
	\cos k_x L = \cos \lambda L \cos P + \sin \lambda L \sin P,
\end{equation}
where $\lambda = E/(\hbar v_F)$. This dispersion relation is 
also periodic in $P$.

\begin{figure}[ht]
  \begin{center}
	\includegraphics[height=3.2cm]{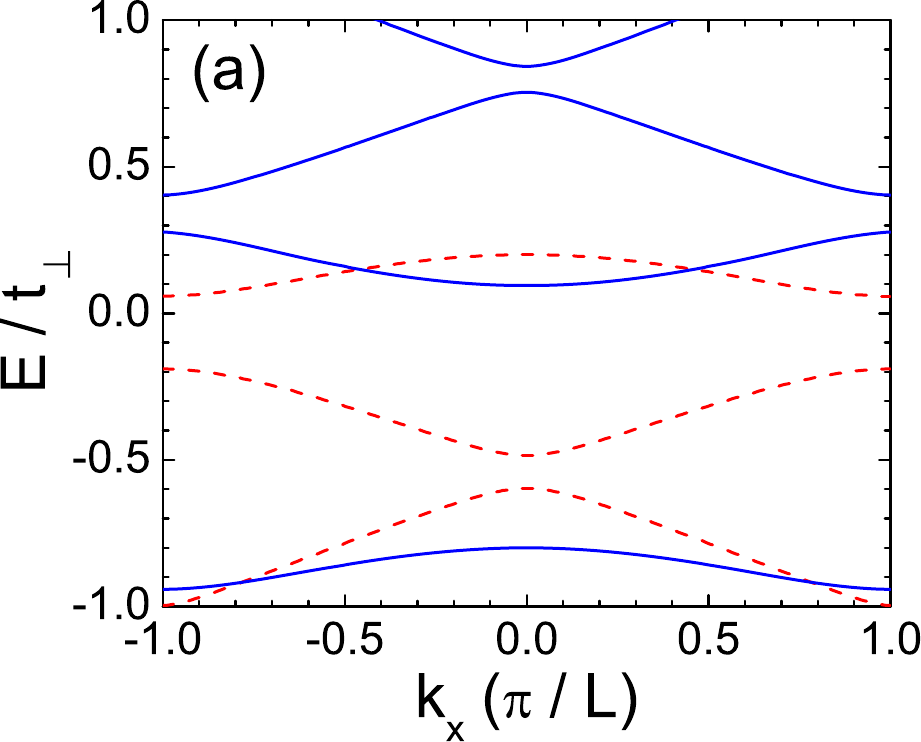}
	\includegraphics[height=3.2cm]{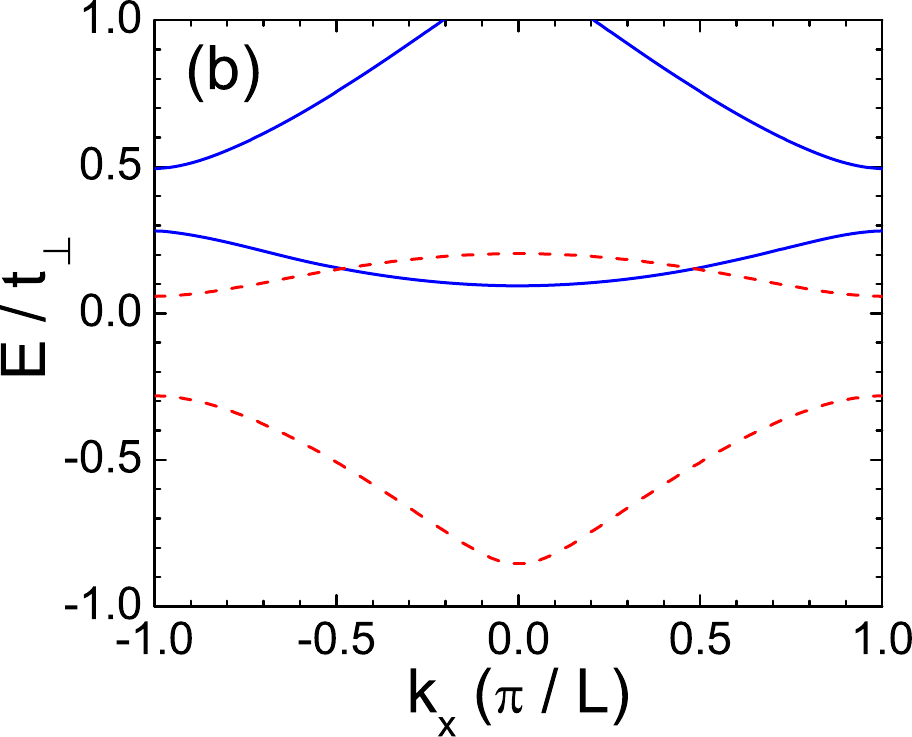}\\
	\includegraphics[height=3.2cm]{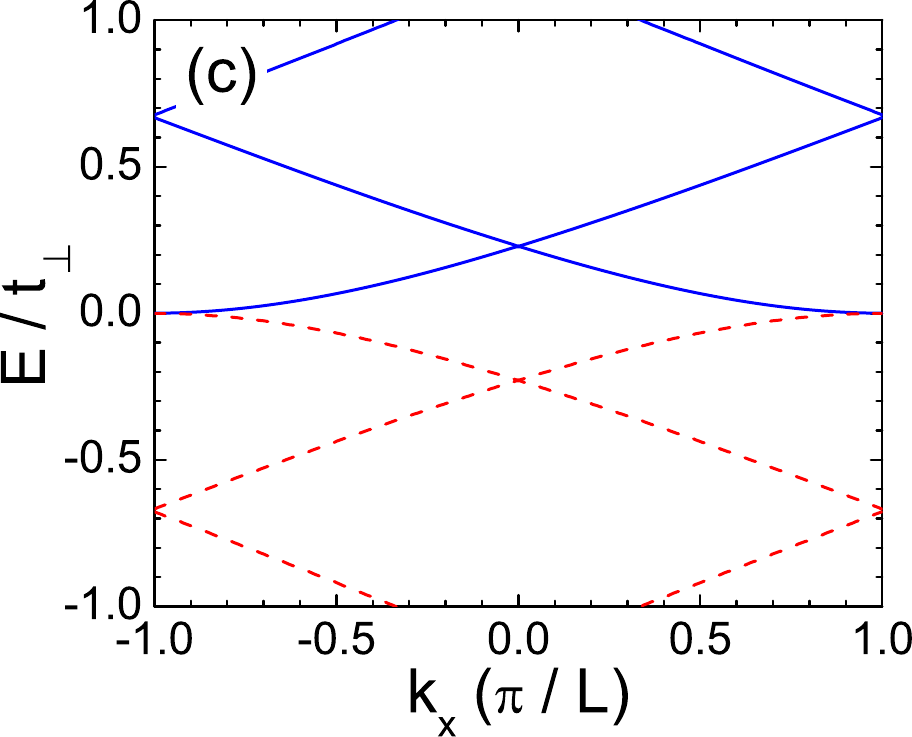}
	\includegraphics[height=3.2cm]{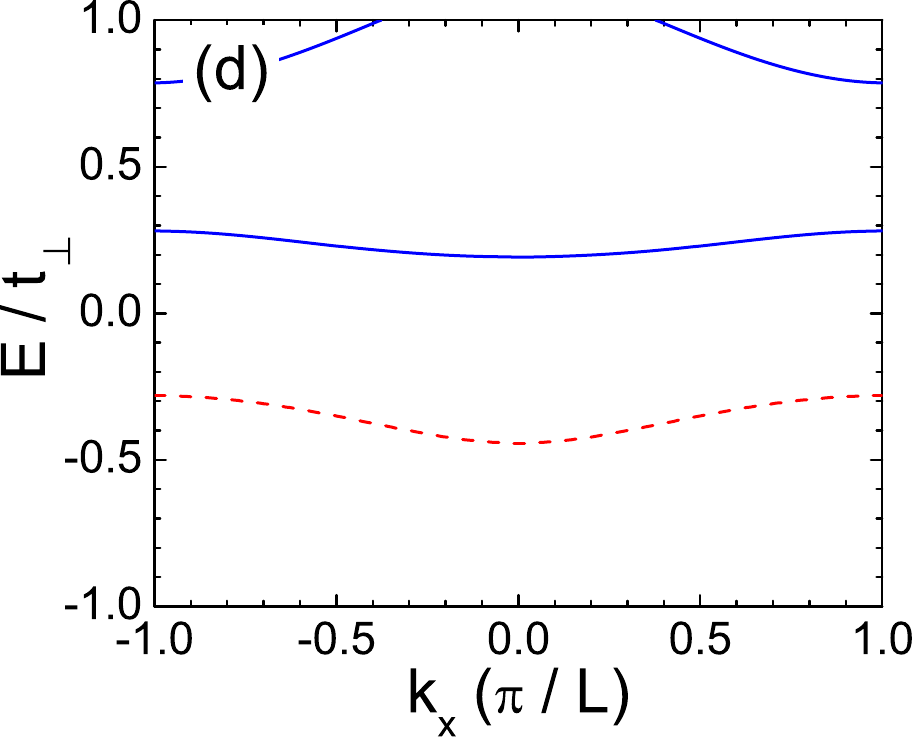}
    \end{center}
  \caption{(Color online) Slices of the spectrum of a KP SL with $L = 10$ $nm$ 
 	along $k_x$, for $k_y = 0$, with $P = 0.25 \pi$ in (a) and  (b) and 
 	$P = \pi$ in (c) and (d). 
  The results in (a) and (c) are obtained from the four-band Hamiltonian (1) and 
  those in (b) and (d) from the two-band one (3). 
  The solid and dashed curves originate, respectively,
   from Eqs. (\ref{eq6_5}a,\ref{eq6_7}a) and (\ref{eq6_5}b,\ref{eq6_7}b).
  }\label{fig7}
\end{figure}

In Fig. \ref{fig7} we plot slices of the energy spectrum for $k_y = 0$. There is 
a qualitative difference, between the four-band and the two-band 
approximation for $P = \pi$. Only when $P$ is small 
does the difference between the 
two 1D dispersion relations become small. Therefore, we will no longer 
present results from the $2\times 2$ Hamiltonian though it has been used 
frequently due to its simplicity. 
The present results indicate that one should be very careful
when using the $2 \times 2$ Hamiltonian in
bilayer graphene.

Notice that for $P = 0.25 \pi$ the electron and hole bands overlap and
cross each other close to $|k_y| \approx 0.5(\pi/L)$. 
That is, this is the spectrum of a semi-metal.
These crossing points move to the edge of the Brillouin zone (BZ) for $P = \pi$ 
resulting in a zero-gap semiconductor.
At the edge of the 
BZ the spectrum is parabolic for low energies.

For $k_y \neq 0$, the dispersion relation can 
be written explicitly in the form
\begin{equation}\label{eq6_8}
	\cos 2 k_x L + C_1 \cos k_x L + C_0/2 = 0,
\end{equation}
where
\begin{equation}\label{eq6_9}
C_1 = -2 (\cos \alpha L + \cos \beta L) \cos P - \left( d_\alpha + d_\beta \right) \sin P,
\end{equation}
and
\begin{eqnarray}\label{eq6_10}
\nonumber
C_0 &=& (2 + k_y^2/\ve^2) + (2 - k_y^2/\ve^2)\cos \alpha L \cos \beta L\\*
\nonumber
&+& [(\ve^2 - k_y^2)^2 + \ve^2(2\ve^2-1)] \sin \alpha L \sin \beta L/2 \alpha \beta \ve^2\\*
\nonumber
&-& 
\Big\{k_y^2/\ve^2 - (2 + k_y^2/\ve^2) \cos \alpha L \cos \beta L \\* 
\nonumber
&+& \left[2\ve^2 - 1/2 - k_y^2 + k_y^4/\ve^2 \right] \sin \alpha L \sin \beta L /\alpha \beta
\Big\} \cos2 P\\*
&+& \left[d_\alpha \cos \beta L + d_\beta \cos \alpha L \right] \sin 2 P,
\end{eqnarray}
with $d_\alpha = (2\ve + 1)\sin \alpha L/\alpha$ and $d_\beta = (2\ve - 1)\sin \beta L/\beta$. 
The wave vectors $\alpha = [\ve^2 + \ve - k_y^2]^{1/2}$ and $\beta= [\ve^2 - \ve - k_y^2]^{1/2}$ are pure real or 
imaginary. If $\beta$ becomes imaginary, the dispersion relation is still real 
($ \beta \rightarrow i |\beta|$ and 
$\sin \beta L \rightarrow i\sinh |\beta| L$). Further, if $\alpha$ becomes 
imaginary,  that is for $\alpha \rightarrow i|\alpha|$, 
the dispersion relation is 
real. The dispersion relation 
has the following invariance properties: 
\begin{subequations}\label{eq6_11}
\begin{eqnarray}  
	 1)\, E(k_x,k_y,P) &=& E(k_x,k_y,P+2n\pi),\\
	 2)\, E(k_x,k_y,P) &=& -E(\pi/L - k_x,k_y,\pi - P),\\
	 3)\, E(k_x,k_y,P) &=& E(k_x,-k_y,P).
\end{eqnarray}
\end{subequations}
\begin{figure}[ht]
  \begin{center}
	\includegraphics[height=3.5cm]{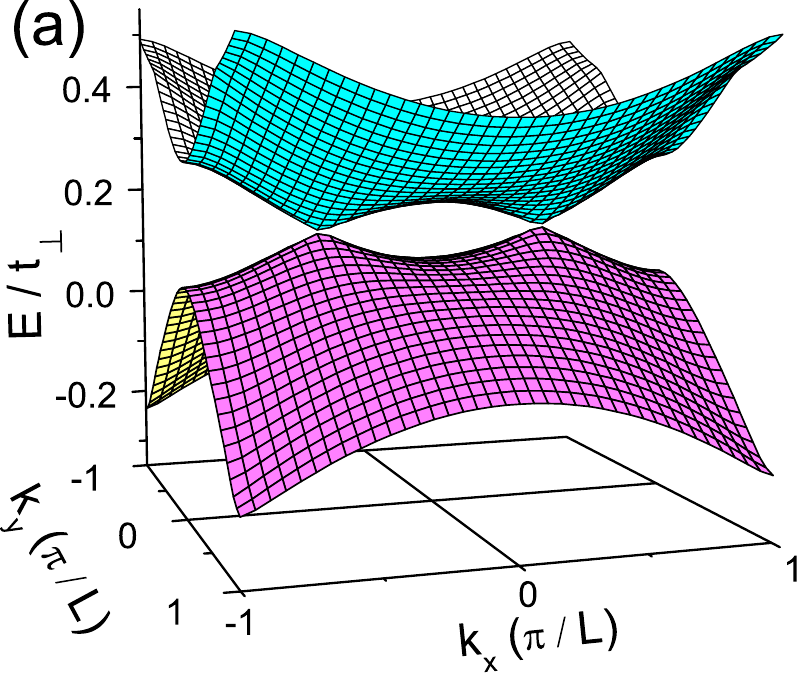}
	\includegraphics[height=3.5cm]{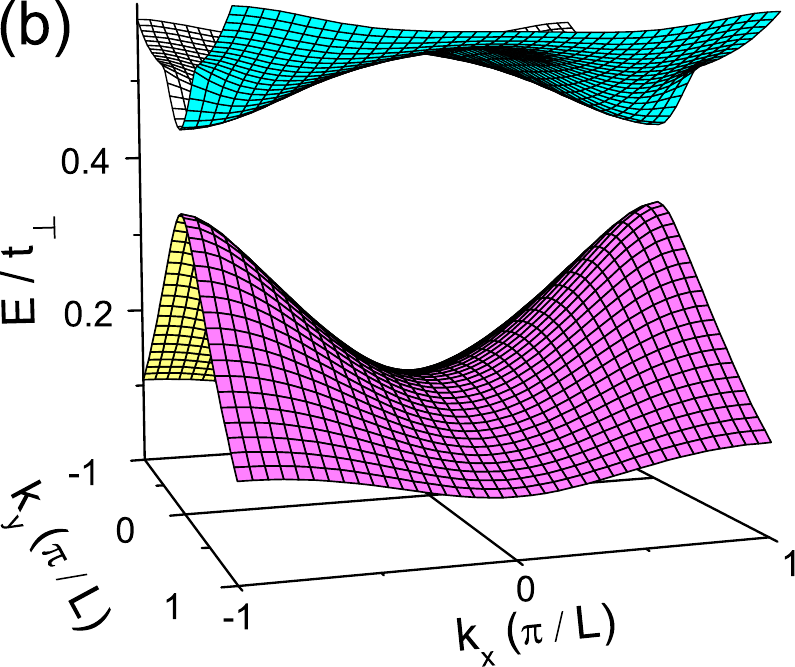}
    \end{center}
  \caption{
  (Color online) SL Spectrum for $L = 10$ nm, the lowest conduction and highest valence band 
  for $P = 0.25 \pi$ in (a) and  $P = 0.5 \pi$ in (b), are shown.}
  \label{fig8}
\end{figure}
\begin{figure}[ht]
  \begin{center}
	\includegraphics[height=3.3cm]{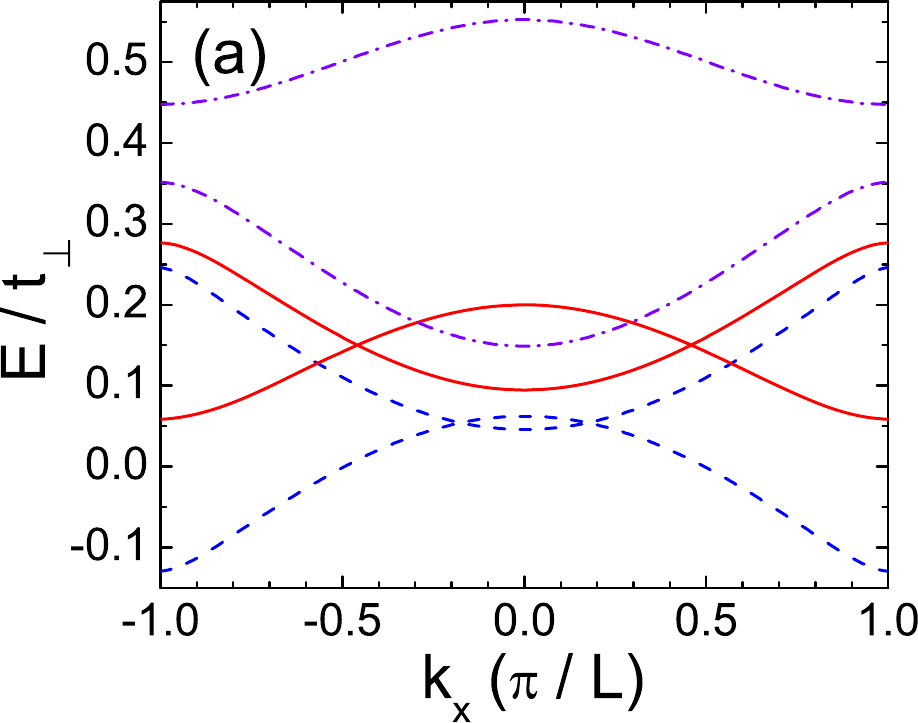}
	\includegraphics[height=3.3cm]{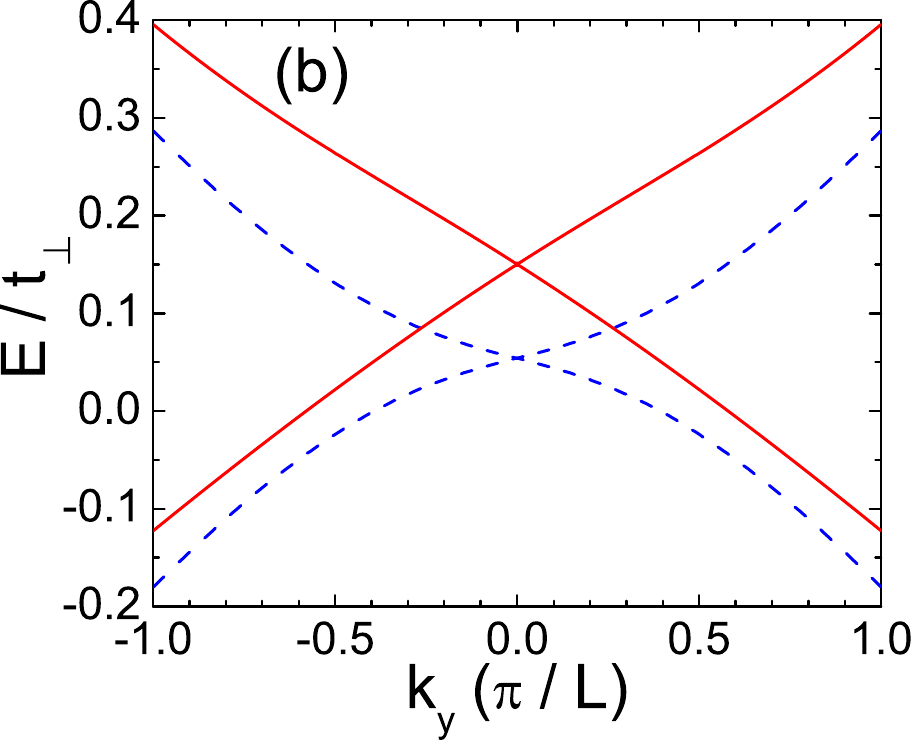}
\vspace*{0.3cm}
	\includegraphics[height=3.3cm]{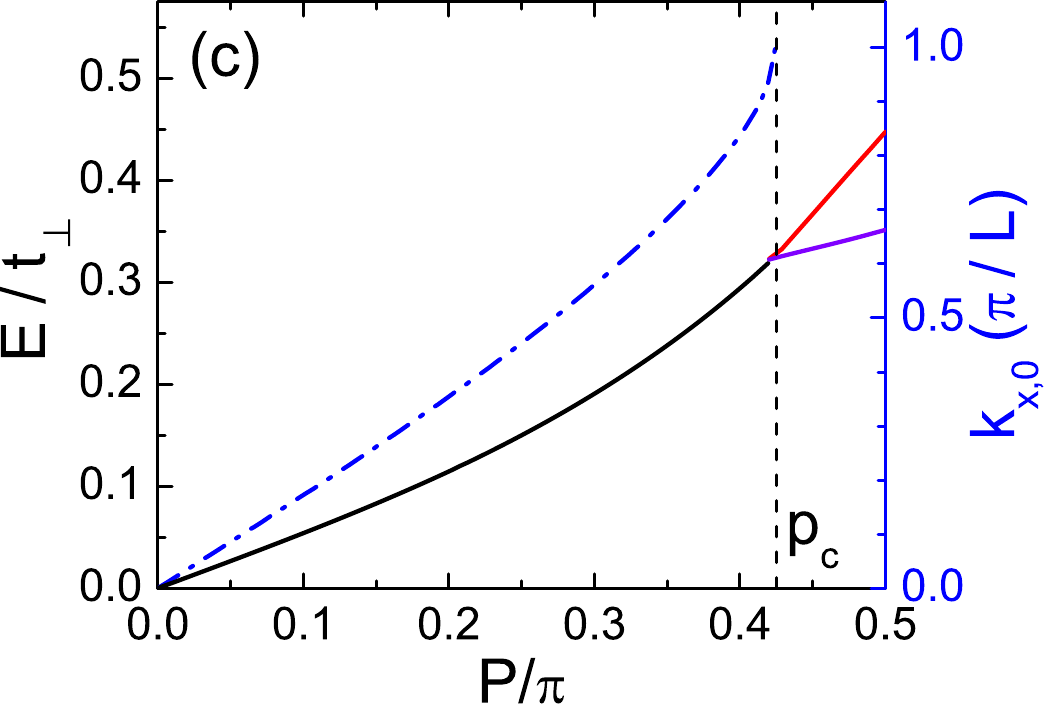}
    \end{center}
  \caption{(Color online) SL spectrum for $L = 10$ $nm$. 
  The dashed blue, solid red, and dash-dotted purple curves are, respectively, 
  for strengths $P = 0.1 \pi$, $P = 0.25 \pi$, and $P = 0.5 \pi$.
  (a) shows the spectrum vs $k_x$ for $k_y = 0$ while (b) shows it 
  vs $k_y$ for $k_x$ at the value where the bands cross. The position of the 
  touching points and the size of the energy gap are shown in (c) as a function 
  of $P$. 
  The dash-dotted, blue curve and the solid, black curve show  $k_{x,0}$ and 
  the energy value of the touching points, respectively. For $P > P_c$ a gap appears; 
 the energies of the conduction band minimum 
  and of the valence band maximum are shown by the red and purple, solid curve, 
  respectively.
 }\label{fig9}
\end{figure}
In Fig. \ref{fig8} we show the lowest conduction and highest valence bands of 
the energy spectrum of the KP model for $P = 0.25\pi$ in (a) and $P = 0.5\pi$ in 
(b). The former has two touching points which can also be viewed as overlapping 
conduction and valence bands as in a semi-metal and the latter exhibits an 
energy gap. 
In Fig. \ref{fig9} slices of Figs. \ref{fig8}(a), (b) 
are plotted for $k_y = 0$. 
The spectrum of bilayer graphene has a single touching point at the origin. 
When the 
strength $P$ is small, this point shifts away from zero energy along the $k_x$ 
axis with $k_y = 0$ and  splits into two points. 
It is interesting to know when and where these touching points emerge. To find 
out 
we observe that at the crossing point both relations 
(\ref{eq6_5}) should be fulfilled. If these two relations are subtracted we 
obtain the transcendental equation
\begin{equation}\label{eq6_12}
	0=(\cos \alpha L - \cos \beta L) \cos P + (1/2)\left(d_\alpha - d_\beta \right)\sin P,
\end{equation}
where $d_\alpha = (2\ve + 1)\sin \alpha L/\alpha$ and 
$d_\beta = (2\ve - 1)\sin \beta L/\beta$. We can solve  Eq. (28) numerically for 
the energy $\ve$. 
For small $P$ and small $L$ this energy can be approximated by $\ve = P/L$. 
Afterwards we can put this solution back into one of the dispersion relations to 
obtain $k_x$.

In Figs. \ref{fig9}(a), (b) we show slices along the $k_x$ axis for $k_y = 0$ 
and along the $k_y$ axis for the $k_x$ value of a touching point, $k_{x,0}$. 
We see that as the touching points move away from the $K$ point, the 
cross sections show a more linear behaviour in the $k_y$ direction. The position 
of the touching points is plotted in Fig. \ref{fig9}(c) as a function of $P$. 
The dash-dotted blue curve corresponds to the value of $k_{x,0}$ 
(right $y$ 
axis), while the energy value of the touching point is given by the 
black solid curve. This touching point moves to the edge of the BZ 
which occurs for  
$P = P_c \approx 0.425 \pi$. 
At this point a gap opens 
(the energies of the top of the valence band 
and of the bottom of the conduction band are shown by the lower purple and upper red solid curve, 
respectively) and 
increases with $P$. Because of property 2) in Eq. \ref{eq6_11}(b) we plot 
the results only for $P < \pi/2$.
We draw attention to the fact that the dispersion relation differs to large 
extent for large $P$ from the one that results 
from the $2\times2$ Hamiltonian given in Appendix \ref{app3}. This is
already apparent from the fact that the dispersion relation does not exhibit 
any of the periodic in $P$ behaviours
given by  Eqs. (\ref{eq6_11}a) and (\ref{eq6_11}b).

\begin{figure*}[ht]
  \begin{center}
	\includegraphics[height=4cm]{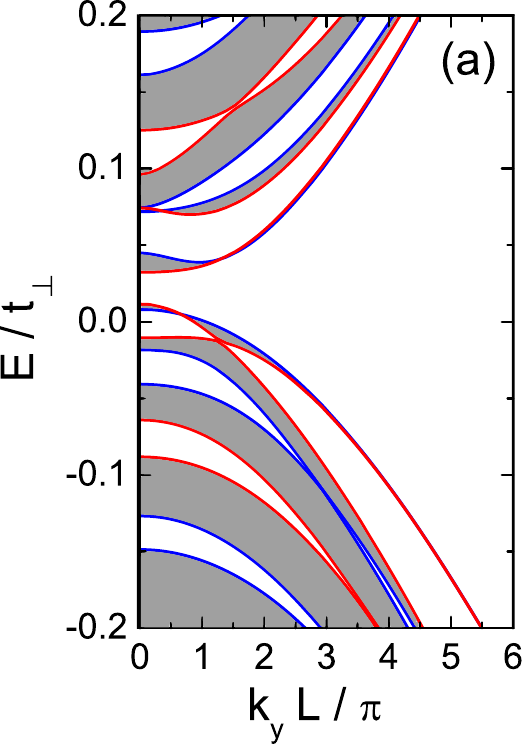}
	\includegraphics[height=4cm]{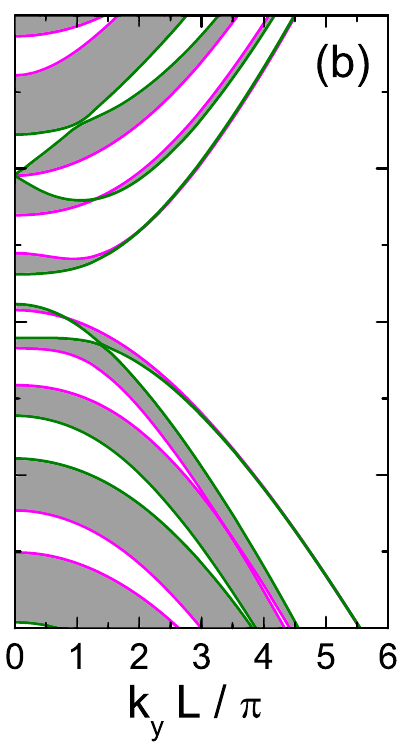}
	\includegraphics[height=4cm]{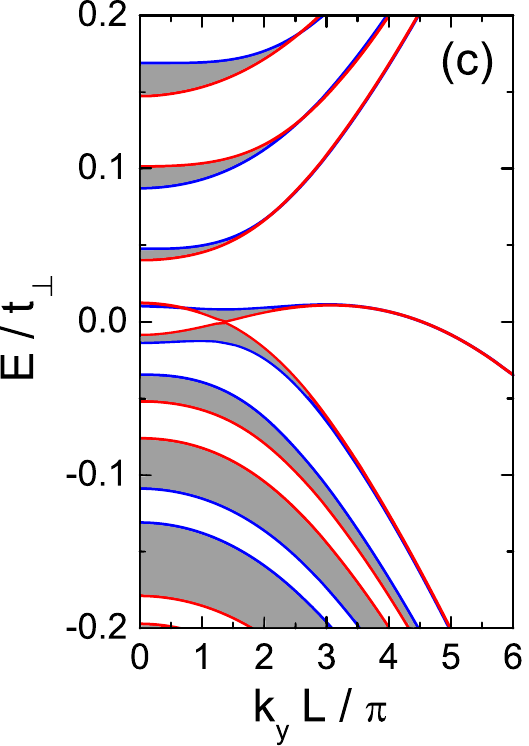}
	\includegraphics[height=4cm]{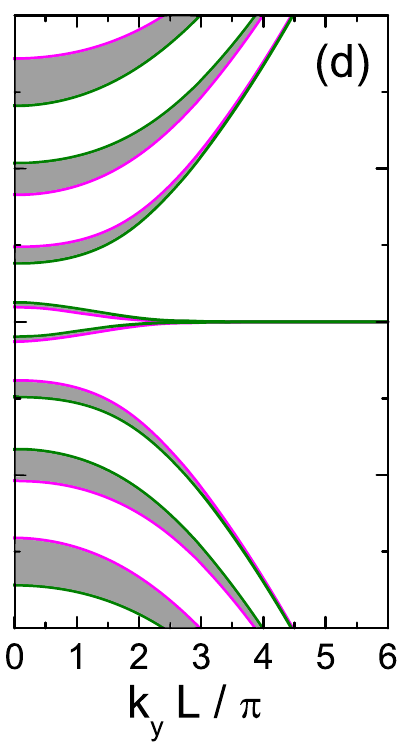}
	\includegraphics[height=3.9cm]{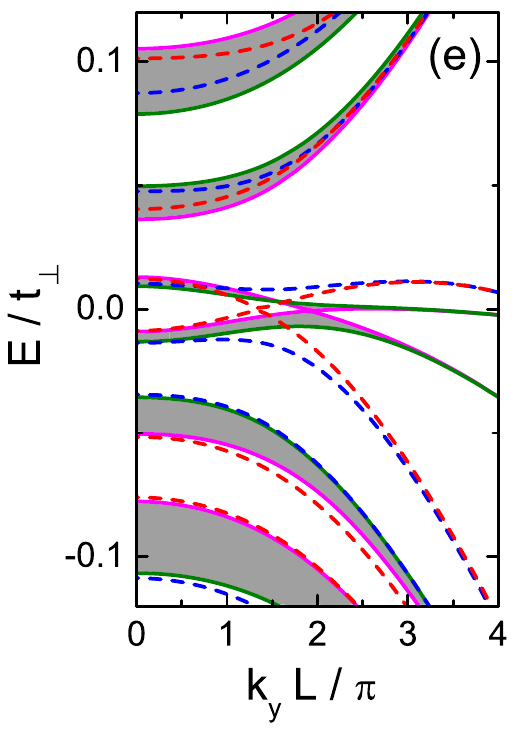}
	\includegraphics[height=4cm]{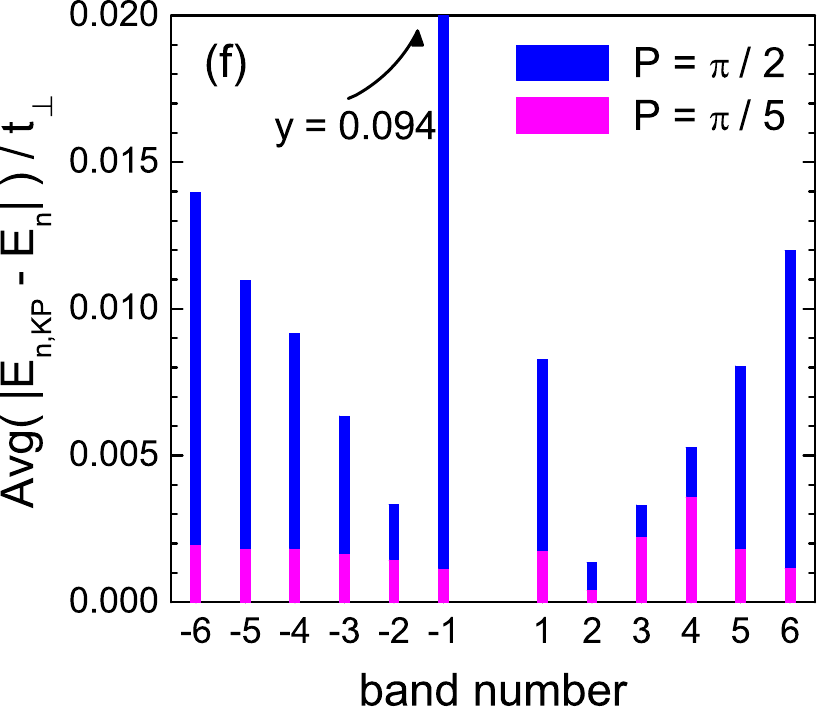}
  \end{center}
  \caption{(Color online) 
  Spectrum of a SL with $L = 50$ nm, 
  (a) and (b) are for $P = 0.2\pi$ and (c) and (d) are for $P = \pi/2$. 
  (a), (c) and (e) are for a rectangular-barrier SL with $W_b = 0.05 L$ and $u = P/W_b$, 
  while (b) and (d) are for the KP model. 
  (e) shows the spectrum for $u$ corresponding to 
  $P = (1/2 + 2)\pi$; 
  the dashed curves show the contours of the spectrum in (c) for $P = \pi/2$.
  (f) Shows the discrepancy of the SL minibands between the exact ones
  and those obtained from the KP model, averaged over {\bf k} space 
  (where we used $k_y L /\pi= 6$ as a cut-off). 
  The conduction (valence) minibands are numbered with positive (negative) integers.
  }\label{fig9extra}
\end{figure*}

An important question is whether the above periodicities in $P$ still remain 
approximately valid outside the range of validity of the KP model. 
To assess that 
we briefly look  at a square-barrier SL 
with barriers of finite width $W_b$ and compare  the 
spectra with those of the KP model. We assume the height of the barrier to be 
$V/\hbar v_F = P/W_b$, such that $V W_b/\hbar v_F = P$.  The SL period 
we use is $50$ nm and the 
width $W_b = 0.05L = 2.5$ nm.
For $P = \pi/2$ the corresponding height is then $V \approx t_\perp$. 
To fit in the continuum model 
we require that the potential barriers be smooth over 
the carbon-carbon distance which is $a \approx 0.14$ nm.
In Fig.~\ref{fig9extra} we show 
the spectra 
for the KP model and this SL. 
Comparing (a) and (b) we see that for $P$ small the 
difference between both models is rather small. If we take $P = \pi/2$ though, 
this difference becomes large, especially for the first conduction and valence minibands,
as shown in panels (c) and (d).
The latter energy bands are flat for large $k_y$ in the KP model, while they 
diverge from the horizontal line (E=0) for a finite barrier width. 
From panel (f), which shows the discrepancy of the SL minibands between the exact ones 
and those obtained from the KP model, we see that the spectra with 
$P = 0.2 \pi$ are closer to the KP model than those for $P = \pi/2$.
Fig.~\ref{fig9extra}(e) demonstrates that the periodicity of the spectrum in 
$P$ within the KP model, i.e., its invariance 
under the change 
$P \rightarrow P + 2n\pi$, is present only as a rough approximation  away from it.
\section{Extended Kronig-Penney model}
In this model we replace the single $\delta$-function barrier in the unit cell 
by two barriers with strengths $P_1$ and $-P_2$. Then the SL potential is 
given by
\begin{equation}\label{eq7_1}
	V(x) = P_1 \sum_n \delta(x-n L) - P_2 \sum_n \delta(x-(n + 1/2) L).
\end{equation}
Here we will restrict ourselves to the important case of $P_1 = P_2$. 
For this potential we can also use Eq. (\ref{eq6_4}) of 
Sec. \ref{sec6}, with 
the transfer matrix $\mathcal{N}$ 
replaced by the appropriate one of Sec. \ref{sec5}.

First, let us consider the spectrum along $k_y = 0$ which is determined by the 
transcendental equations
\begin{subequations}
\begin{eqnarray}\label{eq7_2}
	\cos k_x L  &=& \cos \alpha L \cos^2 P + D_\alpha \sin^2 P ,\\
	\cos k_x L & =& \cos \beta L \cos^2 P + D_\beta \sin^2 P,
\end{eqnarray}
\end{subequations}  
with $D_{\gamma} = \left[(\gamma^2 + \ve^2) \cos \gamma L - \gamma^2 + \ve^2 \right]/4 \gamma^2 \ve^2$. 
It is more convenient to look at the crossing points because 
the spectrum is symmetric around zero energy. 
This follows from the form of the potential (its  spatial average is zero) or 
from the dispersion relation (\ref{eq7_2}): 
the change $\ve \rightarrow -\ve$ entails $\alpha \leftrightarrow \beta$ and 
the crossings in the spectrum are easily obtained by taking the limit 
$\ve \rightarrow 0$ in one of the dispersion relations. This gives 
the value of $k_x$ at the crossings
\begin{equation}\label{eq7_3}
	k_{x,0}  = \pm\arccos[ 1 -(L^2/8) \sin^2 P ]/L,
\end{equation}
and the crossing points are at $(\ve,\,k_x,\,k_y) = 
(0,\,\pm k_{x,0},\,0)$. If 
the $k_{x,0}$ value is not real, then there is no solution at zero energy and a 
gap arises in the spectrum. From Eq. (\ref{eq7_2}) we see that for 
$\sin^2 P > 16/L^2$ a band gap arises.

\begin{figure}[ht]
  \begin{center}
	\includegraphics[height=3.5cm]{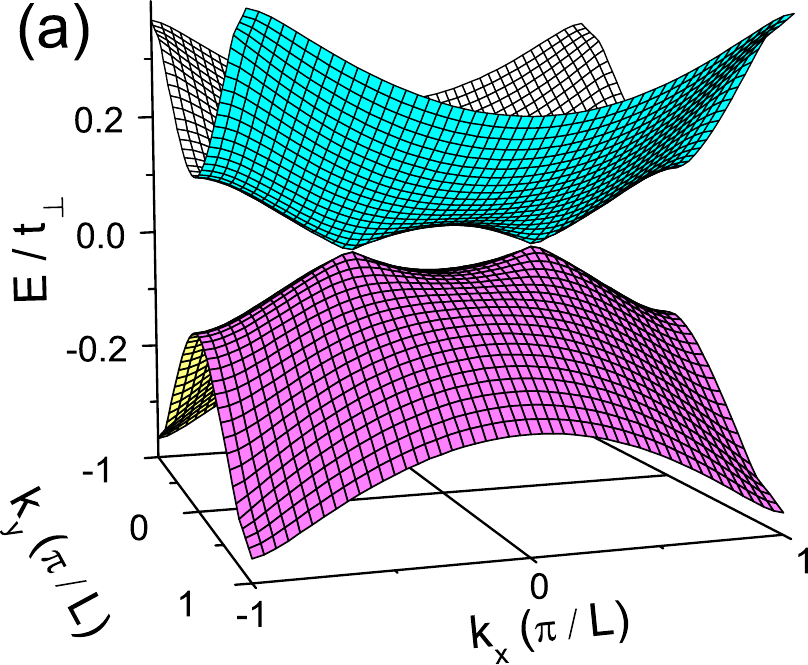}
	\includegraphics[height=3.5cm]{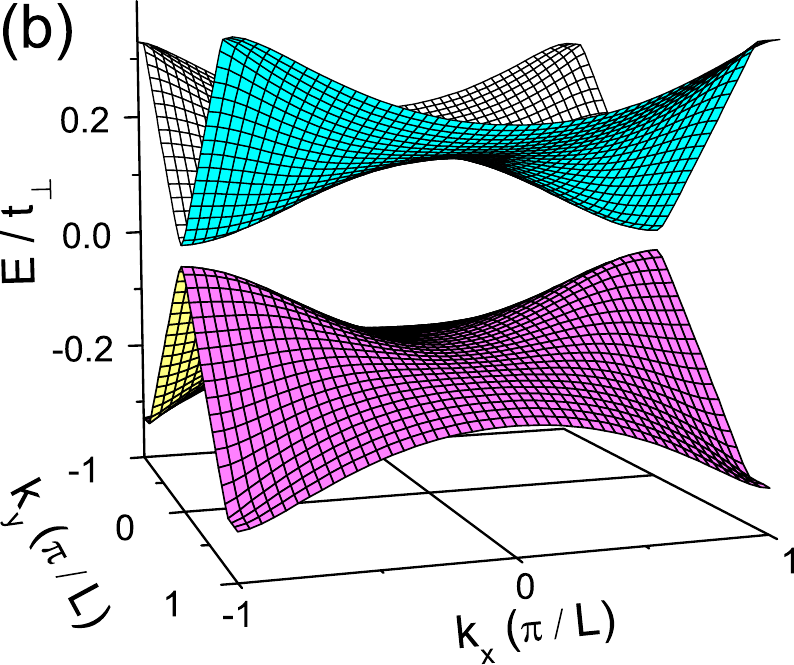}
    \end{center}
  \caption{%
  (Color online) The first conduction and valence 
  minibands for the extended KP 
  model for $L = 10$ nm with $P = 0.125 \pi$ in (a) and $P = 0.25 \pi$ in  (b).}\label{fig10}
\end{figure}
\begin{figure}[ht]
  \begin{center}
	\includegraphics[height=3.5cm]{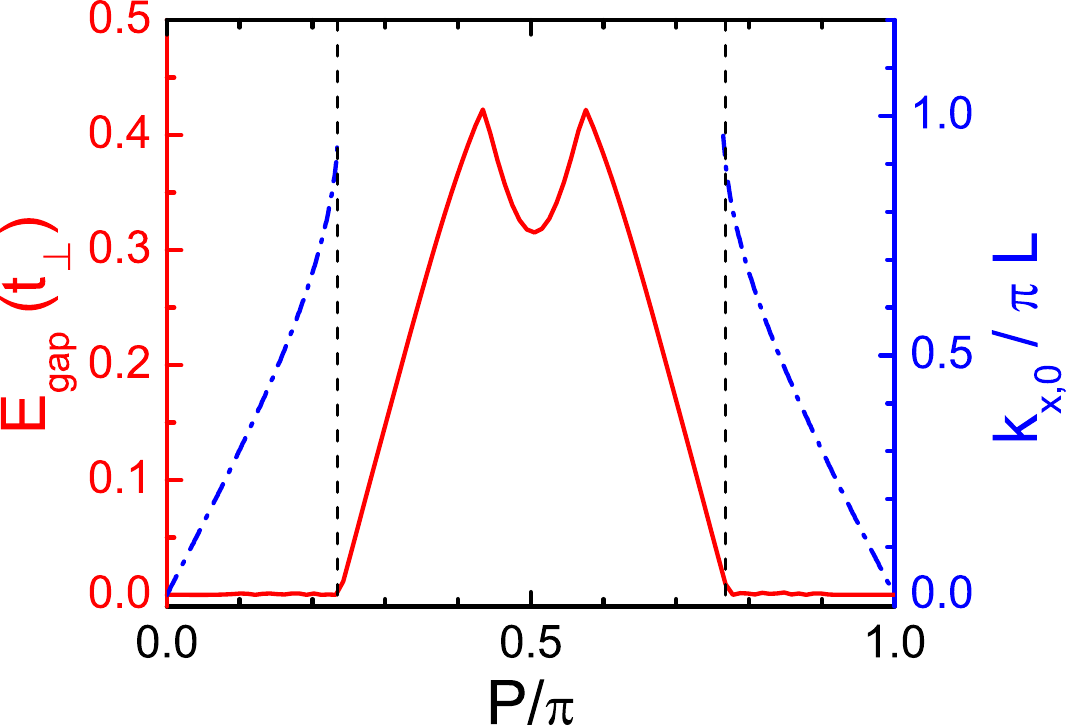}
    \end{center}
  \caption{(Color online) Plot of the  $\pm k_{x,0}$ values, 
  for which the minibands touch each other, 
  as a function of $P$ (dash-dotted, blue curve), and the size of the band gap 
  $E_{gap}$ (solid, red curve). The calculation  is
done for the extended KP model with $L = 10$ nm.}\label{fig11}
\end{figure}

In Fig. \ref{fig10} we show the lowest conduction and 
highest valence band for (a) 
$P = 0.125 \pi$, and (b) $P = 0.25 \pi$. If we make the correspondence with the 
KP model of Sec. V we see that this model leads to qualitatively similar (but 
not identical) spectra shown in
Figs. \ref{fig8}(a) and \ref{fig8}(b): one should 
take $P$ twice as large in the corresponding KP model of Sec. \ref{sec5} in 
order to have a similar 
spectrum.
Here we have the interesting property that the spectrum exhibits mirror symmetry 
with respect to $\ve = 0$ which makes the analysis of the touching points and of 
the gap  easier.

In Fig. \ref{fig11} we plot the   $k_x$ value (dash-dotted, blue curve) of the touching points 
$k_{x,0}$ versus $P$, if there is no gap, and the size of the gap $E_{gap}$ (solid, red curve) if 
there is one. The touching points move toward the BZ boundary with increasing 
$P$. 
Beyond the $P$ value for which  the boundary is reached, 
a gap appears between the conduction and valence 
minibands.
\section{Conclusions}
We investigated the transmission through single and double $\delta$-function 
potential barriers on bilayer graphene using the four-band Hamiltonian. The 
transmission and conductance are found to be {\it  periodic functions of the 
strength of the barriers 
$P=VW_b/\hbar v_F$} with period $\pi$. 
The same periodicity was 
previously 
obtained for such barriers on single-layer graphene \cite{barb3}.
We emphasise 
that the  periodicity obtained here implies that the transmission satisfies the 
relation $T(k_x,k_y,P) = T(k_x,k_y,P + n \pi)$ for arbitrary values of $k_x$, 
$k_y$, $P$, and integer $n$. 
In previous theoretical work on graphene \cite{svi} and bilayer 
graphene \cite{been, mass}  Fabry-P\'erot resonances were studied and 
$T = 1$ was found for particular values of $\alpha$, the electron momentum 
inside the barrier along the $x$ axis. For a rectangular 
barrier of width $W$ and Schr\"odinger-type electrons, Fabry-P\'erot resonances 
occur for $\alpha W = n \pi$ and $E > V_0$ 
as well as in the case of a quantum well for $E > 0$,  $V_0 < 0$. 
In graphene, because of Klein tunnelling, the latter condition on energy is not 
needed. 
Because  $\alpha$ depends on the energy and the potential barrier height in the 
combination $E-V_0$, any periodicity of $T$ in the energy is equivalent to a 
periodicity in $V_0$ 
if no approximations are made, e.g., $E\ll V_0$, etc. 
Although this may appear similar to the 
periodicity in $P$, there are fundamental differences. 
As shown in Ref. \onlinecite{been}, the Fabry-P\'erot resonances are not exactly 
described by the condition $\alpha W = n \pi$ (see Fig.~3 in 
Ref.~\onlinecite{been}) while the periodicity of $T$ in the effective barrier 
strength is exactly $n \pi$. 
Furthermore, the Fabry-P\'erot resonances are found for $T = 1$, while the 
periodicity of $T$ in $P$ is valid for any value of $T$ 
between $0$ and $1$.

Further, we studied  the spectrum of the KP model and found it to be 
{\it periodic in the strength $P$} with period 
$2\pi$. In the extended KP model this period reduces to $\pi$. 
This difference is a consequence of the fact that for the extended SL the unit 
cell contains two $\delta$-function 
barriers. These periodicities are identical to the one found earlier in the 
(extended) KP model on single-layer graphene.
We found that the SL conduction and valence 
minibands touch each other at two points or that there is a 
energy gap between them. In addition, we found a simple relation describing the position of 
these touching points. None of these periodic behaviours results from the 
two-band Hamiltonian; this clearly 
indicates that the two-band Hamiltonian 
is an incorrect description of the KP model in bilayer graphene.
In general, results derived from these two tight-binding Hamiltonians agree 
well only for small energies \cite{cast}. The precise energy ranges are not 
explicitly known and may depend on the particular property studied. 
For the range pertaining to the four-band Hamiltonian ab-initio results \cite{lat} 
indicate that it is approximately from $-1$ eV to + $0.6$ eV.

The question arises whether the above periodicities in $P$ survive when the 
potential barriers have a finite width. To assess that 
we briefly investigated the spectrum of a rectangular SL potential with 
thin barriers and compared it with that 
in the KP limit. We showed with some examples that for specific SL parameters 
the KP model is acceptable in a narrow range of $P$ and only as a rough 
approximation away from this range.
The same conclusion holds for the periodicity of the KP model.

The main differences between the results of this work and those of our previous 
one, Ref. \onlinecite{barb3}, are as follows. In contrast to monolayer graphene 
we found here that:\\
1) The conductance for a single $\delta$-function potential barrier depends on 
the Fermi energy and drops almost to zero for certain values 
 of $E$ and $P$.
2) The KP model (and its extended version) in bilayer graphene can 
open a band gap; if there is no such gap, two touching points appear in the 
spectrum instead of one.
3) The Dirac line found in 
the extended KP model in single-layer graphene is not found in bilayer graphene.
\begin{acknowledgments} This work was supported by IMEC,
the Flemish Science Foundation (FWO-Vl), the Belgian Science Policy
(IAP), and the Canadian NSERC Grant No. OGP0121756.
\end{acknowledgments}
\appendix
\section{Eigenvalues and eigenstates for a constant potential}\label{app1}

Starting with the Hamiltonian 
(\ref{eq2_1}) for a one-dimensional potential $V(x,y) = V(x)$, 
the time-independent Schr\"odinger equation $\mathcal{H} \psi = E \psi$ leads to 
\begin{equation}\label{app1_2}
\begin{aligned}
	- i (\pdje{x} - k_y) \psi_B & = \ve'\psi_A - \psi_{B'},\\
	- i (\pdje{x} + k_y) \psi_A & = \ve'\psi_B, \\
	- i (\pdje{x} + k_y) \psi_{A'} & = \ve'\psi_{B'} - \psi_A,\\
	- i (\pdje{x} - k_y) \psi_{B'} & = \ve'\psi_{A'},
\end{aligned}
\end{equation}

The spectrum and the corresponding eigenstates can be obtained, 
for constant $V(x,y) = V$, 
by progressive elimination of the unknowns in Eq. (\ref{app1_2}) and solution 
of the resulting second-order differential equations. 
The result for the spectrum is
\begin{equation}\label{app1_3}
\begin{aligned}
	\ve & = u + 1/2 \pm \sqrt{1/4 + k^2},\\
	\ve & = u -1/2 \pm \sqrt{1/4 + k^2}.
\end{aligned}
\end{equation}
The unnormalised eigenstates are given by the columns of the matrix	
$\mathcal{G} \mathcal{M}$, where
\begin{equation}\label{app1_4}
	\mathcal{G} = 
		\begin{pmatrix}
			1 & 1 & 1 & 1\\
			f^\alpha_+ & f^\alpha_- & f^\beta_+ & f^\beta_-\\
			-1 & -1 & 1 & 1\\
			f^\alpha_- & f^\alpha_+ & -f^\beta_- & -f^\beta_+\\
		\end{pmatrix},
\end{equation}
with $f^{\alpha,\beta}_\pm = -i(k_y \pm i {(\alpha,\beta)})/\ve'$; 
$\alpha = [\ve'^2 + \ve' - k_y^2]^{1/2}$ and 
$\beta = [\ve'^2 - \ve' - k_y^2]^{1/2}$ are the wave vectors. 
$\mathcal{M}$ is given by
\begin{equation}\label{app1_5}
	\mathcal{M} = 
		\begin{pmatrix}
			e^{i \alpha x} & 0 & 0 & 0\\
			0 & e^{- i \alpha x} & 0 & 0\\
			0 & 0 & e^{i \beta x} & 0\\
			0 & 0 & 0 & e^{-i \beta x}\\
		\end{pmatrix}.
\end{equation}
The wave function in a region of constant potential is a linear combination 
of the eigenstates and can be written
\begin{equation}\label{app1_6}
	\Psi(x) = \kvecc{\psi_{A}}{\psi_{B}}{\psi_{B'}}{\psi_{A'}} = \mathcal{G} \mathcal{M} \kvecc{A}{B}{C}{D}.
\end{equation}
We can reduce its complexity by the linear transformation
 $\Psi(x)\to \mathcal{R}\Psi(x)$ where
\begin{equation}\label{app1_7}
	\mathcal{R} = \frac{1}{2}
		\begin{pmatrix}
			1 & 0 & -1 & 0\\
			0 & 1 & 0 & -1\\
			1 & 0 & 1 & 0\\
			0 & 1 & 0 & 1\\
		\end{pmatrix},
\end{equation}
which transforms $\Psi(x)$ to $\Psi(x) = (1/2) (\psi_{A} - \psi_{B'}, \, \psi_{B} - \psi_{A'}, \, \psi_{A} + \psi_{B'}, \, \psi_{B} + \psi_{A'})^T$. Then the basis functions are given by the columns of $\mathcal{G} \mathcal{M}$ with
\begin{equation}\label{app1_8}
	\mathcal{G} = 
		\begin{pmatrix}
			1 & 1 & 0 & 0\\
			\alpha/\ve' & -\alpha/\ve' & -ik_y/\ve' & -ik_y/\ve'\\
			0 & 0 & 1 & 1\\
			-ik_y/\ve' & -ik_y/\ve' & \beta/\ve' & -\beta/\ve'\\
		\end{pmatrix}.
\end{equation}
The matrix $\mathcal{M}$ is unchanged under the transformation $\mathcal{R}$ 
and the new $\Psi(x)$ fulfils the same boundary conditions as the old one. 

\section{The transfer matrix}\label{app2}

We denote the wave function to the left of, inside, and to the right of the barrier by 
$\psi_j(x) = \mathcal{G}_j\mathcal{M}_j\mathcal{A}_j$, with 
$j = 1$, $2$, and $3$,  respectively. Further, we have $\mathcal{G}_1 = \mathcal{G}_3$ and 
$\mathcal{M}_1 = \mathcal{M}_3$. The continuity of the wave function 
at $x = 0$ and $x = W_b$  gives the boundary conditions $\psi_1(0) = \psi_2(0)$ 
and $\psi_2(W_b) = \psi_3(W_b)$. 
In explicit matrix notation this gives 
$\mathcal{G}_1\mathcal{A}_1 = \mathcal{G}_2\mathcal{A}_2$ and 
$\mathcal{G}_2\mathcal{M}_2(W_b)\mathcal{A}_2 = \mathcal{M}_1(W_b) \mathcal{G}_1\mathcal{A}_3$, 
where $\mathcal{A}_1 = \mathcal{G}_1^{-1} \mathcal{G}_2 \mathcal{M}^{-1}_2(W_b) \mathcal{G}^{-1}_2 \mathcal{G}_1 \mathcal{M}_1(W_b) \mathcal{A}_3$.
Then the transfer matrix $\mathcal{N}$ can  be written as 
$\mathcal{N} = \mathcal{G}_1^{-1} \mathcal{G}_2 \mathcal{M}^{-1}_2(W_b) \mathcal{G}^{-1}_2 \mathcal{G}_1 \mathcal{M}_1(W_b)$. 
Let us define $\mathcal{N}' = \mathcal{G}_2 \mathcal{M}^{-1}_2(W_b) \mathcal{G}^{-1}_2$, 
which leads to $\psi_1(0) = \mathcal{N}' \psi_3(W_b)$.

To treat the case of a $\delta$-function barrier we take the limits 
$V \rightarrow \infty$ and $W_b \rightarrow 0$ such that the dimensionless 
potential strength $P = V W_b / \hbar v_F$ is kept constant. 
Then $\mathcal{G}_2$ and  $\mathcal{M}_2(W_b)$ simplify to 
\begin{equation}\label{app2_4}
	\mathcal{G}_2 = 
	\begin{pmatrix}
		1 & 1 & 0 & 0\\
		-1 & 1 & 0 & 0\\
		0 & 0 & 1 & 1\\
		0 & 0 & -1 & 1
	\end{pmatrix},
\end{equation}
\begin{equation}\label{app2_5}
	\mathcal{M}_2(W_b) = 
	\begin{pmatrix}
		e^{i P} & 0 & 0 & 0\\
		0 & e^{-i P} & 0 & 0\\
		0 & 0 & e^{i P} & 0\\
		0 & 0 & 0 & e^{-i P}
	\end{pmatrix},
\end{equation}
and $\mathcal{N}'$ becomes
\begin{equation}\label{app2_6}
	\mathcal{N}' = 
	\begin{pmatrix}
		\cos P & i \sin P & 0 & 0\\
		i \sin P & \cos P & 0 & 0\\
		0 & 0 & \cos P & i \sin P\\
		0 & 0 & i \sin P & \cos P
	\end{pmatrix}.
\end{equation}

\section{Results for the $2\times2$ Hamiltonian}\label{app3}

Using the $2\times2$ Hamiltonian (\ref{eq2_3}) instead of the $4\times4$ 
one can  sometimes 
lead to unexpectedly different results; below we give a few examples. 
In a slightly modified notation pertinent to 
the $2\times2$ Hamiltonian 
we set $\alpha = [-\ve + k_y^2]^{1/2}$, $\beta = [\ve + k_y^2]^{1/2}$, 
and use the same dimensionless units as before.

Bound states for a single $\delta$-function barrier $u(x) = P\delta(x)$, 
without accompanying propagating states, are possible if $k_y = 0$ or $k_y^2 > |\ve|$. 
In the former case the single solution is $\ve = -sign(P) P^2/4$. 
In the latter one the dispersion relation is 
\begin{equation}\label{app3_1}
	\ve^2 (P + 2 \alpha) (P - 2 \beta) + 2 P^2 k_y^2 (\alpha \beta - k_y^2) = 0.
\end{equation}

The dispersion relation for the KP model obtained from the $2\times2$ Hamiltonian is
\begin{equation}\label{app3_2}
	\cos(2 k L) + 2 F_1 \cos(k L) + F_2 = 0,
\end{equation}
where
\begin{equation}\label{app3_3}
\begin{aligned}
	&F_1 = -\cosh(\beta L) - \cosh(\alpha L) + \frac{P}{2\beta}\sinh(\beta L) - \frac{P}{2\alpha}\sinh(\alpha L),\\
	&F_2 = \frac{1}{\alpha \beta \ve^2} \left\{\alpha \beta (\ve^2 + k_y^2 P^2/4)\right. \\
	& \quad + \beta \cosh(\beta L) \left[ \alpha (2 \ve^2 - k_y^2)\cosh(\alpha L) + \ve^2 P \sinh(\alpha L) \right]\\
	& \left.\quad - \frac{P}{2} \sinh(\beta L) \left[ \alpha (\ve^2 - k_y^4/2)P \sinh(\alpha L) + 2 \ve^2 \alpha \cosh(\alpha L) \right] \right\}.
\end{aligned}
\end{equation}

\end{document}